\documentclass[a4paper,fleqn,usenatbib,useAMS]{mnras}
\usepackage{graphicx}	
\usepackage{amsmath}	
\usepackage{amssymb}	
\usepackage{multicol}        
\usepackage{bm}		
\usepackage{pdflscape}	

\usepackage[T1]{fontenc}
\usepackage{ae,aecompl}
\usepackage{comment}

\newcommand{\arctanh}{\mathop{\mathrm{arctanh}}\nolimits}

\newcommand{\lr}[1]{\left( #1 \right)}

\newcommand{\lrt}[1]{\left< #1 \right>}
\newcommand{\lrv}[1]{\left| #1 \right|}

\newcommand{\bal}{{\boldsymbol \alpha}}
\newcommand{\bbe}{{\boldsymbol \beta}}
\newcommand{\bpa}{{\boldsymbol \partial}}
\newcommand{\bth}{{\boldsymbol \theta}}

\newcommand{\bga}{{\boldsymbol \gamma}}

\newcommand{\mbg}{{\mbox{\boldmath$g$}}}

\newcommand{\mbF}{{\mbox{\boldmath$F$}}}
\newcommand{\mbG}{{\mbox{\boldmath$G$}}}

\newcommand{\cC}{{\cal C}}

\newcommand{\cE}{{\cal E}}
\newcommand{\cF}{{\cal F}}
\newcommand{\cG}{{\cal G}}

\newcommand{\cI}{{\cal I}}
\newcommand{\cL}{{\cal L}}

\newcommand{\cZ}{{\cal Z}}

\newcommand{\mbcF}{{\mbox{\boldmath$\cF$}}}
\newcommand{\mbcG}{{\mbox{\boldmath$\cG$}}}

\newcommand{\tbbe}{{\tilde{\bbe}}}

\newcommand{\convergence}{{\kappa}}
\newcommand{\shear}{{\bga}}
\newcommand{\flexionfirst}{{\mbF}}
\newcommand{\flexionsecond}{{\mbG}}

\newcommand{\reducedshear}{{\mbg}}
\newcommand{\reducedflexionfirst}{{\mbcF}}
\newcommand{\reducedflexionsecond}{{\mbcG}}

\newcommand{\lensingdistortion}[2]{{L^{#1}_{#2}}}

\newcommand{\lensingreduceddistortion}[2]{{\cL^{#1}_{#2}}}

\newcommand{\intrinsicdistortion}[2]{{\cI^{#1}_{#2}}}

\newcommand{\combineddistortion}[2]{{\cC^{#1}_{#2}}}

\newcommand{\eigendistortion}[2]{{\cE^{#1}_{#2}}}

\newcommand{\eigencdistortion}[2]{{\cE^{'#1}_{#2}}}

\newcommand{\cnv}{{\convergence}}
\newcommand{\sh}{{\shear}}
\newcommand{\flf}{{\flexionfirst}}
\newcommand{\fls}{{\flexionsecond}}

\newcommand{\ldst}[2]{{\lensingdistortion{#1}{#2}}}

\newcommand{\lrdst}[2]{{\lensingreduceddistortion{#1}{#2}}}

\newcommand{\idst}[2]{{\intrinsicdistortion{#1}{#2}}}

\newcommand{\rsh}{{\reducedshear}}
\newcommand{\rflf}{{\reducedflexionfirst}}
\newcommand{\rfls}{{\reducedflexionsecond}}

\newcommand{\cdst}[2]{{\combineddistortion{#1}{#2}}}

\newcommand{\edst}[2]{{\eigendistortion{#1}{#2}}}

\newcommand{\ecdst}[2]{{\eigencdistortion{#1}{#2}}}

\newcommand{\WGLMD}{\Omega}
\newcommand{\zmom}[2]{{\cZ^{#1}_{#2}}}
\newcommand{\object}{S}

\usepackage{newtxtext,newtxmath}

\title{The impact of higher-order distortions on the precise measurement of weak gravitational lensing shear and flexion.}

\author[Yuki Okura, Toshifumi Futamase]
{Yuki Okura,$^{1,2}$\thanks{E-mail: yuki.okura@nao.ac.jp}
Toshifumi Futamase,$^{3}$
\\
$^{1}$NAOJ\\
$^{2}$RIKEN\\
$^{3}$Tohoku University}
\begin{document}
\label{firstpage}
\pagerange{\pageref{firstpage}--\pageref{lastpage}}
\maketitle

\begin{abstract}
In this paper, we investigate the impact of higher-order distortions on the precise measurement of weak gravitational lensing shear and flexion.
We begin by defining generalized higher-order distortions and outlining methods for measuring them. Then, using several lens models, we examine how these distortions affect shear and flexion measurements.
Our results show that neglecting higher-order distortions can introduce systematic errors of a few percent in both shear and flexion measurements, indicating that these effects cannot be ignored.
Although the strength of these errors depends on factors such as lensing strength and the size of background sources, we demonstrate that simultaneous measurement of higher-order distortions can reduce the systematic errors to below 1\% in most cases.
\end{abstract}
\onecolumn
\section{Introduction}

The importance of high-precision measurements of weak gravitational lensing shear and flexion has been increasing in recent years driven by recent advances in observational equipment. 
For example, in observing cosmic shear to study detailed nature of dark energy, highly accurate shape measurement of background galaxies is required. 
However, recent high resolution observations using space telescope such as JWST and Euclid  have given us huge number of faint galaxies with irregular shapes. Most of these faint galaxies are not directly used in the conventional methodology of weak lensing analysis. 
This will mean that we are loosing some of the lensing information contained in these faint images.  
Therefore it is urgent to establish a new method of shape measurement to fully make use of new information brought from new observations.

For this purpose we have developed a new scheme of weak lensing analysis(see Okura and Futamase 2022) based on the moment method.  
Conventional shear measurements basically link the spin-2 component of the quadrupole moment of the background galaxy shape to gravitational shear, but many of the galaxy shapes images have higher-order multipole moment components. 
These higher-order multipole moments correspond to certain combinations of higher derivatives of the gravitational lensing potential. 
While the shear is the spin-2 component of the second derivative of the potential, the lensing effect also includes the spin-2 component of higher order derivatives of the gravitational potential. Therefore, if measurements are not performed taking these higher-order effects into account, “shear + the higher-order spin-2 component” will be measured as shear, resulting in errors.

Furthermore, flexions, which corresponds to the the spin-1 and spin-3 components of the third derivative of the lens potential, contain important information on the structure of the gravitational potential, which cannot be determined by shear alone, but its measurements are also likely to be subject to errors due to the influence of higher-order effects. In this way, evaluating the influence of higher-order effects in shear and flexion measurements is extremely important in order to extract accurate lens information.
In this paper, we focus on this point, and use a simple lens model to calculate higher-order effects to accurately evaluate their impact on shear and flexion measurements. For recent and general reviews of weak lensing, we ask reader to consult for example Mandelbaum 2018 and Futamase and Okura 2025. 

This paper is organized as follows.
Section 2 summarizes the definitions and notations used throughout the paper.
Section 3 introduces the generalized weak gravitational lensing distortions and describes the methods used to measure the distortions.
Section 4 presents theoretical calculations of higher-order lensing effects for several lens models.
Section 5 examines the systematic errors that arise depending on whether higher-order lensing effects are taken into account, using multiple lens models.
Finally, Section 6 provides the conclusions.
\section{Notations and Definitions}
In this section, we summarize the basic formulations used throughly in this paper. 
We adopt complex notation for coordinates, lens equations, and other quantities relevant to weak lensing analysis.
\subsection{Coordinates and Deviation}
We begin by defining the coordinate systems and deviations used in this work.

The coordinates of the celestial sphere on the image plane are defined in complex form as
\begin{eqnarray}
\label{eq:coordinates}
\bth = \theta_1 + i\theta_2,
\end{eqnarray}
and an infinitesimal angular displacement from the center of the background object, denoted as $\overline{\bth}$, is expressed as
\begin{eqnarray}
\label{eq:infinitesimalangulardisplacement }
d\bth = \bth - \overline\bth.
\end{eqnarray}
Similarly to the above definitions, we denote the source-plane coordinate as $\bbe$, and the zero-plane coordinate as $\tbbe$(for the zero-plane we will explain more detail in the below.)

To describe weak lensing distortions, we use a complex differential operator defined as
\begin{eqnarray}
\label{eq:deviations}
\nabla \equiv \lr{\frac{\partial}{\partial\theta_1} +i\frac{\partial}{\partial\theta_2}}
\end{eqnarray}
This operator satisfies $\nabla \bth = 0$ and $\nabla^* \bth = 2$, where $^*$ denotes complex conjugation. 
Note that this operator differs from the standard complex derivative in mathematical complex analysis.

We also define a complex partial derivative operator as
\begin{eqnarray}
\label{eq:deviations_complex}
\bpa \equiv \frac{\partial}{\partial\bth} = \frac12\nabla^*
\end{eqnarray}
which satisfies $\bpa \bth = 1$ and $\bpa \bth^* = 0$.

\subsection{Lens equation and distortions}
The lens equation is described as follows 
\begin{eqnarray}
\label{eq:lensequation}
\bbe = \bth - \bal = \bth - \nabla\psi
\end{eqnarray}
where $\bal$ is the deflection angle by lensing effect and it can be defined as a deviation of the lensing potential $\psi$. 
Then the expansion of the equation \ref{eq:lensequation} until 3rd order deviation is described as follows
\begin{eqnarray}
\label{eq:lensedquation}
d\bbe &=& 
d\bth - \cnv d\bth - \sh d\bth^*-\frac14\lr{\flf^*d\bth^2_2 +2\flf d\bth^{2}_0+\fls d\bth^2_{-2}} 
\nonumber\\&=&
\lr{1-\cnv}\lr{d\bth  - \rsh d\bth^*-\frac14\lr{\rflf^*d\bth^2_2 +2\rflf d\bth^{2}_0+\rfls d\bth^2_{-2}} }
\end{eqnarray}
where $\cnv$, $\sh$, $\flf$ and $\fls$ are weak gravitational lensing convergence, shear, first flexion and second flexion defined as the second and third deviations of the lens potential $\psi$
\begin{eqnarray}
\label{eq:convergence}
\cnv &\equiv& \frac12\nabla\nabla^*\psi
\\
\label{eq:shear}
\sh &\equiv& \frac12\nabla\nabla\psi
\\
\label{eq:flexion1}
\flf &\equiv& \frac12\nabla\nabla\nabla^*\psi
\\
\label{eq:flexion2}
\fls &\equiv& \frac12\nabla\nabla\nabla\psi
\end{eqnarray}
and $\rsh$, $\rflf$ and $\rfls$ are the weak gravitational lensing reduced shear, reduced first flexion and reduced second flexion, and
\begin{eqnarray}
\label{eq:dtNM}
d\bth^N_M = d\bth^Pd\bth^{*Q}
\end{eqnarray}
with $P=\frac{N+M}2$ and $Q=\frac{N-M}2$.

\section{The Generalized weak gravitational distortions}
In this section, we present the definitions of generalized weak lensing distortions, and introduce the method for measuring the generalized distortions. 

\subsection{The Generalized weak gravitational lensing distortions}
The generalized forms of \textit{the weak gravitational lensing distortions} $\ldst{N}{M}$ are defined as follows:
\begin{eqnarray}
\label{eq:DNM}
\ldst{N}{M} \equiv \frac12\nabla^N_M \psi
\end{eqnarray}
where the generalized differential operator is defined as
\begin{eqnarray}
\label{eq:NNM}
\nabla^N_M \equiv \nabla^{P} \nabla^{*Q}.
\end{eqnarray}
Hereafter, the weak lensing distortions - potential, deflection angle, convergence, shear, and flexions - are denoted in the generalized notation as: $\ldst{0}{0} = \psi/2$, $\ldst{1}{1} = \bal/2$, $\ldst{2}{0} = \cnv$, $\ldst{2}{2} = \sh$, $\ldst{3}{1} = \flf$ and $\ldst{3}{3} = \fls$.

By using the definitions, equation \ref{eq:lensedquation} can be extended to generalized forms as follows:
\begin{eqnarray}
\label{eq:lensedquation_multi}
d\bbe &=& d\bth - \lr{\ldst{2}{0} d\bth + \ldst{2}{2} d\bth^*}
\nonumber\\&&
-\frac14\lr{\ldst{3}{-1}d\bth^2_2 +2\ldst{3}{1} d\bth^{2}_0+\ldst{3}{3} d\bth^2_{-2}} 
\nonumber\\&&
-\frac1{24}\lr{{\ldst{4}{-2}}d\bth^3_3 + 3\ldst{4}{0}d\bth^3_1 +3\ldst{4}{2}d\bth^3_{-1} + \ldst{4}{4}d\bth^3_{-3}} 
\nonumber\\&&
-\frac1{192}\lr{{\ldst{5}{-3}}d\bth^4_4 + 4{\ldst{5}{-1}}d\bth^4_2 + 6\ldst{5}{1}d\bth^4_0 + 4\ldst{5}{3}d\bth^4_{-2} + \ldst{5}{5}d\bth^4_{-4}} + \cdots
\\&=&
\lr{1-\ldst{2}{0}}\Biggl(d\bth  - \lrdst{2}{2} d\bth^*
\nonumber\\&&
-\frac14\lr{\lrdst{3}{-1}d\bth^2_2 +2\lrdst{3}{1} d\bth^{2}_0+\lrdst{3}{3} d\bth^2_{-2}} 
\nonumber\\&&
-\frac1{24}\lr{{\lrdst{4}{-2}}d\bth^3_3 + 3\lrdst{4}{0}d\bth^3_1 +3\lrdst{4}{2}d\bth^3_{-1} + \lrdst{4}{4}d\bth^3_{-3}} 
\nonumber\\&&
-\frac1{192}\lr{{\lrdst{5}{-3}}d\bth^4_4 + 4{\lrdst{5}{-1}}d\bth^4_2 + 6\lrdst{5}{1}d\bth^4_0 + 4\lrdst{5}{3}d\bth^4_{-2} + \lrdst{5}{5}d\bth^4_{-4}} + \cdots
\Biggr)
\\&=&
\lr{1-\ldst{2}{0}}\lr{d\bth - \sum_{N=1}^{\infty}\sum_{k=0}^N{\binom{N}{2k-N}}'\lrdst{N+1}{N+1-2k}d\bth^N_{2k-N}}
\\&\equiv&
\lr{1-\ldst{2}{0}}\lr{d\bth - \WGLMD(\lrdst{}{}, d\bth)}
\end{eqnarray}
where \textit{the weak gravitational lensing reduced distortions} $\lrdst{N}{M}$ are defined as follows:
\begin{eqnarray}
\label{eq:rDNM}
\lrdst{N}{M} \equiv \frac{\ldst{N}{M}}{1-\ldst{2}{0}} \hspace{10pt} (2 \leqq N).
\end{eqnarray}
and modified binomial coefficients ${\binom{N}{M}}'$ 
in \textit{the expansion terms of the weak gravitational lensing equation} $\WGLMD$
is defined
\begin{eqnarray}
\label{eq:CNM}
{\binom{N}{M}}' \equiv
\begin{cases}
  0 & (N=2,\, M=0), \\[6pt]
  \binom{N}{M} & \text{otherwise}.
\end{cases}
\end{eqnarray}

\subsection{The Generalized Intrinsic and Combined Distortions}
Background galaxies have their own intrinsic shapes, and the observed images are further distorted by the lensing effect. Since the intrinsic shapes cannot be separated from the lensing distortions, they behave as intrinsic noise.
In this work, we adopt the concept of \textit{the zero plane} introduced in Okura and Futamase (2016), treating the intrinsic shapes as intrinsic distortions. In other words, the intrinsic shape of a background galaxy is considered to result from a transformation applied to a circular source in the zero plane $\tbbe$.
For the latest developments regarding the zero plane, see Okura and Futamase (2022).
The relationship between the zero plane and the source plane can be expressed as follows.
\begin{eqnarray}
\label{eq:lensedquation_multi_I}
d\tbbe &=& d\bbe - \WGLMD(\idst{}{}, d\bbe)
\end{eqnarray}
where $\idst{N}{M}$ means \textit{the weak gravitational intrinsic distortions}, 

Finally, the observed shape of a background object is the result of the combination of intrinsic distortion and lensing distortion, referred to as \textit{the weak gravitational combined distortions} $\cdst{N}{M}$, and is expressed as follows:
\begin{eqnarray}
\label{eq:lensedquation_multi_C}
d\tbbe &=& \lr{1-\ldst{2}{0}}\lr{1 + \idst{2}{2}\lrdst{2}{-2}}\lr{d\bth - \WGLMD(\cdst{}{}, d\bth)}
\end{eqnarray}

We show some of the weak gravitational combined distortion explicitly for later convenience. 
\begin{eqnarray}
\label{eq:rshear_C}
\cdst{2}{2} &=& \frac{\idst{2}{2} + \lrdst{2}{2}}{1 + \idst{2}{2}\lrdst{2}{-2}}\\
\label{eq:rflexion1_C}
\cdst{3}{1} &=& \frac{\idst{3}{1} + \lrdst{3}{1} - \idst{2}{2}\lrdst{3}{-1} - \lrdst{2}{2}\idst{3}{-1}- \lrdst{2}{-2}\idst{3}{3} + \lrv{\lrdst{2}{2}}^2\idst{3}{1}}{1 + \idst{2}{2}\lrdst{2}{-2}}\\
\label{eq:rflexion2_C}
\cdst{3}{3} &=& \frac{\idst{3}{3} + \lrdst{3}{3} - \idst{2}{2}\lrdst{3}{1} - 2\lrdst{2}{2}\idst{3}{1}+ \lr{\lrdst{2}{2}}^2\idst{3}{-1}}{1 + \idst{2}{2}\lrdst{2}{-2}}\\
\label{eq:rflexion3_C}
\cdst{3}{-1} &=& \frac{\idst{3}{-1} + \lrdst{3}{-1} - \idst{2}{2}\lrdst{3}{-3} - 2\lrdst{2}{-2}\idst{3}{1} + \lr{\lrdst{2}{-2}}^{2}\idst{3}{3}}{1 + \idst{2}{2}\lrdst{2}{-2}}\\
\label{eq:rD4B_C}
\cdst{4}{-2}&=& \frac{\idst{4}{-2}+ \lrdst{4}{-2}- \idst{2}{2}\lrdst{4}{-4}- 3\lrdst{2}{2}c\idst{4}{0}+ 3\lr{\lrdst{2}{-2}}^{2}\idst{4}{2}- \lr{\lrdst{2}{-2}}^{3}\idst{4}{4}}{1 + \idst{2}{2}\lrdst{2}{-2}}\\
\label{eq:rD40_C}
\cdst{4}{0} &=& \frac{\idst{4}{0} + \lrdst{4}{0} - \idst{2}{2}\lrdst{4}{-2}- \lrdst{2}{2}\idst{4}{-2}+ 2\lrv{\lrdst{2}{2}}^2\idst{4}{0}- \lrdst{2}{-2}\lr{2+\lrv{\lrdst{2}{2}}^2}\idst{4}{2}+ \lr{\lrdst{2}{-2}}^{2}\idst{4}{4}}{1 + \idst{2}{2}\lrdst{2}{-2}}\\
\label{eq:rD42_C}
\cdst{4}{2} &=& \frac{\idst{4}{2} + \lrdst{4}{2} - \idst{2}{2}\lrdst{4}{0}+ \lr{\lrdst{2}{2}}^2\idst{4}{-2}- \lrdst{2}{2}\lr{2+\lrv{\lrdst{2}{2}}^2}\idst{4}{0}+ 2\lrv{\lrdst{2}{2}}^2\idst{4}{2}- \lrdst{2}{-2}\idst{4}{4}}{1 + \idst{2}{2}\lrdst{2}{-2}}\\
\label{eq:rD44_C}
\cdst{4}{4} &=& \frac{\idst{4}{4} + \lrdst{4}{4} - \idst{2}{2}\lrdst{4}{2}- \lr{\lrdst{2}{2}}^3\idst{4}{-2}+ 3\lr{\lrdst{2}{2}}^2\idst{4}{0}- 3\lrdst{2}{2}\idst{4}{2}}{1 + \idst{2}{2}\lrdst{2}{-2}},
\end{eqnarray}
where the terms involving the square of the flexion in $\cdst{4}{X}$ are omitted. 
The expressions for $\cdst{5}{X}$ are also omitted, as they are too lengthy and complex to be presented here. These quantities should be computed numerically within the analysis code.
Here, it is important to note that, unlike lensing distortions or intrinsic distortions, the combined distortion does not satisfy the correspondence between negative spin numbers and complex conjugation. In other words, except in special cases, one has $\cdst{3}{1} \neq \lr{\cdst{3}{-1}}^*$ , $\cdst{4}{-2}\neq \lr{\cdst{4}{2}}^*$ and this discrepancy persists for higher-order moments as well.

The term $\lr{1-\ldst{2}{0}}\lr{1 + \idst{2}{2}\lrdst{2}{-2}}$ represents the change in size and rotation of the source image in the source plane, and since it is not meaningful for weak lensing analyses, it will be neglected hereafter.

\subsection{The Generalized Distortions Measurement Method: GEM}
\label{sec:measurement}
In this section, we introduce the method for measuring the generalized combined distortions defined in the previous section. 
The measurement technique for the combined flexions was introduced by Okura and Futamase (2022), which involves defining zero moments and identifying the lensing distortions that make these moments value zero. Since the measurement of generalized distortions is a straightforward extension of this idea to higher-order moments, this paper focuses only on presenting the definitions of the zero moments, and refers the reader to Okura and Futamase (2022) for the detailed idea and calculations.

The zero moments $\zmom{N}{M}$ of the background object image $\tilde \object(\tbbe) = \object(\bth)$ is defined in the zero plane as follows 
\begin{eqnarray}
\zmom{N}{M} &\equiv& \int d^2\tilde\beta d\tbbe^M_N \tilde \object(\tbbe) \tilde W (|d\tbbe|) = \int d^2\theta J(\bth) \lr{d\bth - \WGLMD(\cdst{}{}, d\bth)}^M_N \object(\bth) W(d\bth)\\
d\tbbe^M_N&=&d\tbbe^Pd\tbbe^{*Q}\\
J(\bth) &\equiv& \lrv{\frac{d^2\tilde\beta}{d^2\theta}},
\end{eqnarray}
where $\tilde W(|d\tbbe|)$ is a weight function for reducing noise from random count and $J(\bth)$ is Jacobian.
By identifying the combined distortions $\cdst{N}{M}$ such that all zero moments except for the spin-0 component are set to zero, the combined distortions can be measured. In this paper, we refer to this measurement scheme for generalized distortions as \textit{the GEneralized distortions Measurement method: GEM.} Since it is not feasible in practice to measure distortions up to infinite order, it is necessary to determine the maximum order to be measured according to the purpose of the analysis. In this work, we denote the case where distortions are measured up to the N th order as GEM-N th. For example, GEM-2nd indicates that the measurement has been performed up to the order of shear.

Here, since the number of zero moments that must be set to zero is smaller than the number of parameters associated with the centroid and combined distortions, there exist infinitely many combinations of combined distortions that satisfy the zero-moment condition.

\subsection{The Eigen Distortions}
As discussed in previous sections, the combined distortions possess degrees of freedom that satisfy the zero-moment condition and are therefore not uniquely determined. 

As the first step to solve this problem, we introduce \textit{the weak gravitational eigen distortions} $\edst{N}{M}$. These quantities are independent of these degrees of freedom and can be obtained by making use of the fact that, in the zero plane, the image is circular, i.e., a function of $|d\tbbe|^2$.
The eigen distortions were originally defined in Okura and Futamase (2022), but here we redefine them in a more general form to account for the degrees of freedom associated with $\cdst{2}{2}$, and extends until 5th order distortions as follows.
\begin{eqnarray}
\label{eq:rconvergence_E}
\edst{2}{0} &\equiv& \frac{2}{1+\lrv{\cdst{2}{2}}^2}\lr{-2\cdst{2}{2}\cdst{2*}{2}}\\
\label{eq:rshear_E}
\edst{2}{2} &\equiv& \frac{2}{1+\lrv{\cdst{2}{2}}^2}\cdst{2}{2}\\
\label{eq:rflexion1_E}
\edst{3}{1} &\equiv& \frac2{1+\lrv{\cdst{2}{2}}^2}\lr{2\cdst{3}{1} + \cdst{3}{-1}^* - 2\cdst{2}{2}\cdst{3}{1}^* - \cdst{2}{2}^*\cdst{3}{3}}\\
\label{eq:rflexion2_E}
\edst{3}{3} &\equiv& \frac2{1+\lrv{\cdst{2}{2}}^2}\lr{\cdst{3}{3} - \cdst{2}{2}\cdst{3}{-1}^*}\\
\label{eq:rD40_E}
\edst{4}{0} &\equiv& \frac{2}{1+\lrv{\cdst{2}{2}}^2}\lr{3\cdst{4}{0} -3\cdst{2}{2}\cdst{4*}{2}}\\
\label{eq:rD42_E}
\edst{4}{2} &\equiv& \frac{2}{1+\lrv{\cdst{2}{2}}^2}\lr{3\cdst{4}{2} +\cdst{4*}{-2} -3\cdst{2}{2}\cdst{4*}{0} -\cdst{2}{2}^*\cdst{4}{4}}\\
\label{eq:rD44_E}
\edst{4}{4} &\equiv& \frac{2}{1+\lrv{\cdst{2}{2}}^2}\lr{\cdst{4}{4} -\cdst{2}{2}\cdst{4*}{-2}}\\
\label{eq:rD51_E}
\edst{5}{1} &\equiv& \frac{2}{1+\lrv{\cdst{2}{2}}^2}\lr{6\cdst{5}{1} + 4\cdst{5*}{-1} -6\cdst{2}{2}\cdst{5*}{1} -4\cdst{2}{2}^*\cdst{5}{3}}\\
\label{eq:rD53_E}
\edst{5}{3} &\equiv& \frac{2}{1+\lrv{\cdst{2}{2}}^2}\lr{4\cdst{5}{3} + \cdst{5*}{-3} -4\cdst{2}{2}\cdst{5*}{-1} -\cdst{2}{2}^*\cdst{5}{5}}\\
\label{eq:rD55_E}
\edst{5}{5} &\equiv& \frac{2}{1+\lrv{\cdst{2}{2}}^2}\lr{\cdst{5}{5} -\cdst{2}{2}\cdst{5*}{-3}},
\end{eqnarray}
where the combined distortions satisfy $\edst{N}{-M} = \edst{N*}{M}$.
Here, to consider an example of such degrees of freedom in the case of reduced shear. 
Let $\cdst{2}{2m}$ be a measured combined shear which satisfies the zero moment condition, its reciprocal of the complex conjugate $1/\cdst{2*}{2m}$ also satisfies the same condition. As a result, it is not possible at this step to determine which of the two measured values are true value of $\cdst{2}{2}$, i.e. $\cdst{2}{2} = \cdst{2}{2m}$ or $\cdst{2}{2} = 1/\cdst{2*}{2m}$?. 
In this sense, the measurable quantity can be reinterpreted as the eigen shear $\edst{2}{2} = 2\cdst{2}{2}/(1+|\cdst{2}{2}|^2)$. This form corresponds to the well-known complex distortion discussed in various previous studies. Hereafter, we consider only the weak lensing regime ($|\cdst{2}{2}| < 1$).

Similarly, for the case of flexion and higher-order distortions, there exist infinitely many combinations of combined distortions that satisfy the zero moment condition. However, regardless of the specific combination, the values of the eigen distortions remain invariant.
Therefore, the true combined distortions can be obtained from the obtained combined distortions, which are merely one of the infinitely many valid solutions, through the eigen distortions.

While the eigen distortions allow us to obtain the combined effect of the intrinsic and lensing distortions, an additional step is required to isolate the lensing distortions from the eigen distortions. Specifically, this involves statistically averaging out the intrinsic distortions, which behave as random noise within the eigen distortions. However, since the eigen flexions include terms proportional to the square of absolute of the random noise i.e. $|\idst{2}{2}|^2$, a straightforward averaging make a systematic bias originating from the intrinsic noise. 
Therefore, it is necessary to reformulate the eigen distortions in a form that excludes such squared noise terms before performing statistical averaging.

\textit{The weak gravitational eigen distortion combinations} $\ecdst{N}{M}$ are also redefined as follows
\begin{eqnarray}
\label{eq:rshear_EC}
\ecdst{2}{2} &\equiv&\frac12\frac{1+\lrv{\cdst{2}{2}}^2}{1-\lrv{\cdst{2}{2}}^2}\lr{\edst{2}{2}+\cdst{2}{2}\edst{2}{0}+\lr{\cdst{2}{2}}^2\edst{2}{2}^*}=\cdst{2}{2}\\
\label{eq:rflexion1_EC}
\ecdst{3}{1} &\equiv&\frac12\frac{1+\lrv{\cdst{2}{2}}^2}{1-\lrv{\cdst{2}{2}}^2}\lr{\edst{3}{1}+\cdst{2}{2}^*\edst{3}{3}+\cdst{2}{2}\edst{3}{1}^*+\lr{\cdst{2}{2}}^2\edst{3}{3}^*}=2\cdst{3}{1}+\cdst{3}{-1}^*+\cdst{2}{2}\cdst{3}{-1}\\
\label{eq:rflexion2_EC}
\ecdst{3}{3} &\equiv&\frac12\frac{1+\lrv{\cdst{2}{2}}^2}{1-\lrv{\cdst{2}{2}}^2}\lr{\edst{3}{3}+\cdst{2}{2}\edst{3}{1}+\lr{\cdst{2}{2}}^2\edst{3}{1}^*+\lr{\cdst{2}{2}}^3\edst{3}{3}^*}=\cdst{3}{3}+2\cdst{2}{2}\cdst{3}{1}+\lr{\cdst{2}{2}}^2\cdst{3}{-1}\\
\label{eq:rD40_EC}
\ecdst{4}{0} &\equiv& \frac12\frac{1+\lrv{\cdst{2}{2}}^2}{1-\lrv{\cdst{2}{2}}^2}\lr{\edst{4}{0} + \cdst{2}{2}\edst{4*}{2} + \lr{\cdst{2}{2}}^2\edst{4*}{4}} = 3\cdst{4}{0} + \cdst{2}{2}\cdst{4}{-2}\\
\label{eq:rD42_EC}
\ecdst{4}{2} &\equiv& \frac12\frac{1+\lrv{\cdst{2}{2}}^2}{1-\lrv{\cdst{2}{2}}^2}\lr{ \edst{4}{2} + \cdst{2}{2}\edst{4*}{0} + \cdst{2}{2}^*\edst{4}{4}} = 3\cdst{4}{2} + \cdst{4*}{-2} \\
\label{eq:rD44_EC}
\ecdst{4}{4} &\equiv&\frac12\frac{1+\lrv{\cdst{2}{2}}^2}{1-\lrv{\cdst{2}{2}}^2}\lr{\edst{4}{4} + \cdst{2}{2}\edst{4}{2} + \lr{\cdst{2}{2}}^2\edst{4}{0}} = \cdst{4}{4} + 3\cdst{2}{2}\cdst{4}{2} \\
\label{eq:rD51_EC}
\ecdst{5}{1} &\equiv& \frac12\frac{1+\lrv{\cdst{2}{2}}^2}{1-\lrv{\cdst{2}{2}}^2}\lr{\edst{5}{1} + \cdst{2}{2}\edst{5*}{1} + \cdst{2}{2}^*\edst{5}{3} + \lr{\cdst{2}{2}}^{*2}\edst{5}{5} + \lr{\cdst{2}{2}}^2\edst{5*}{3} + \lr{\cdst{2}{2}}^3\edst{5*}{5}}
\nonumber\\&&
 = 6\cdst{5}{1} + 4\cdst{5*}{-1} + \cdst{2}{2}^*\cdst{5*}{-3} + 4\cdst{2}{2}\cdst{5}{-1} + \lr{\cdst{2}{2}}^2\cdst{5}{-3} \\
\label{eq:rD53_EC}
\ecdst{5}{3} &\equiv&\frac12\frac{1+\lrv{\cdst{2}{2}}^2}{1-\lrv{\cdst{2}{2}}^2}\lr{\edst{5}{3} + \cdst{2}{2}^*\edst{5}{5} + \cdst{2}{2}\edst{5}{1} + \lr{\cdst{2}{2}}^2\edst{5*}{1} + \lr{\cdst{2}{2}}^3\edst{5*}{3} + \lr{\cdst{2}{2}}^4\edst{5*}{5}}
\nonumber\\&&
 = 4\cdst{5}{3} + \cdst{5*}{-3} + 6\cdst{2}{2}\cdst{5}{1} + 4\lr{\cdst{2}{2}}^2\cdst{5}{-1} + \lr{\cdst{2}{2}}^3\cdst{5}{-3} \\
\label{eq:rD55_EC}
\ecdst{5}{5} &\equiv& \frac12\frac{1+\lrv{\cdst{2}{2}}^2}{1-\lrv{\cdst{2}{2}}^2}\lr{\edst{5}{5} + \cdst{2}{2}\edst{5}{3} + \lr{\cdst{2}{2}}^2\edst{5}{1} + \lr{\cdst{2}{2}}^3\edst{5*}{1} + \lr{\cdst{2}{2}}^4\edst{5*}{3} + \lr{\cdst{2}{2}}^5\edst{5*}{5}}
\nonumber\\&&
 = \cdst{5}{5} + 4\cdst{2}{2}\cdst{5}{3} + 6\lr{\cdst{2}{2}}^2\cdst{5}{1} + 4\lr{\cdst{2}{2}}^3\cdst{5}{-1} + \lr{\cdst{2}{2}}^4\cdst{5}{-3},
\end{eqnarray}
Here, true combined shear is required to obtain the eigen distortion combinations.

The relations between the lensing distortions and the eigen distortion combinations can be obtained by taking statistical average as follows
\begin{eqnarray}
\label{eq:rshear_EC_ave}
\lrt{\ecdst{2}{2}} &=&\lrdst{2}{2}\\
\label{eq:rflexion1_EC_ave}
\lrt{\ecdst{3}{1}} &=&3\lrdst{3}{1}+\lrdst{2}{2}\lrdst{3}{-1}\\
\label{eq:rflexion2_EC_ave}
\lrt{\ecdst{3}{3}} &=&\lrdst{3}{3}+2\lrdst{2}{2}\lrdst{3}{1}+\lr{\lrdst{2}{2}}^2\lrdst{3}{-1}\\
\label{eq:rD40_EC_ave}
\lrt{\ecdst{4}{0}} &=& 3\lrdst{4}{0} + \lrdst{2}{2}\lrdst{4}{-2}\\
\label{eq:rD42_EC_ave}
\lrt{\ecdst{4}{2}} &=& 4\lrdst{4}{2} \\
\label{eq:rD44_EC_ave}
\lrt{\ecdst{4}{4}} &=& \lrdst{4}{4} + 3\lrdst{2}{2}\lrdst{4}{2} \\
\label{eq:rD51_EC}
\lrt{\ecdst{5}{1}} &=& 10\lrdst{5}{1} + \lrdst{2}{-2}\lrdst{5}{3} + 4\lrdst{2}{2}\lrdst{5}{-1} + \lr{\lrdst{2}{2}}^2\lrdst{5}{-3} \\
\label{eq:rD53_EC}
\lrt{\ecdst{5}{3}} &=& 5\lrdst{5}{3} + 6\lrdst{2}{2}\lrdst{5}{1} + 4\lr{\lrdst{2}{2}}^2\lrdst{5}{-1} + \lr{\lrdst{2}{2}}^3\lrdst{5}{-3} \\
\label{eq:rD55_EC}
\lrt{\ecdst{5}{5}} &=& \lrdst{5}{5} + 4\lrdst{2}{2}\lrdst{5}{3} + 6\lr{\lrdst{2}{2}}^2\lrdst{5}{1} + 4\lr{\lrdst{2}{2}}^3\lrdst{5}{-1} + \lr{\lrdst{2}{2}}^4\lrdst{5}{-3}.
\end{eqnarray}
Finally, lensing distortions $\lrdst{N}{M}$ can be obtained by solving the above set of equations simultaneously using the statistical average of the eigen distortion combinations.

\section{Weak Gravitational Lensing Distortions in Lens Models}
\label{sec:lensmodel}
The convergence and shear produced by different lens models have been discussed in various weak lensing studies.
In particular, Meneghetti (2021) provides detailed calculations of weak lensing shear for several lens models.
Additionally, the computation of flexion is thoroughly presented in Bacon et al. (2006).

The following physical parameters are used in this section: 
$r$ is the three-dimensional (3D) radial distance from the center of the lens object, 
$\rho(r)$ is mass distribution of the lens model, 
$\Sigma(\theta)$ is the surface mass density, defined by projecting the three-dimensional mass density $\rho(r)$ along the line of sight (i.e., the $z$-axis). That is,
\begin{eqnarray}
\label{eq:surfacemassdensity}
\Sigma(\theta) = \int dz\, \rho(r)
\end{eqnarray}
then the surface mass density is related to the weak lensing convergence $\ldst{2}{0}$ as follows: 
\begin{eqnarray}
\label{eq:k_surfacemassdensity}
\ldst{2}{0} = \frac{\Sigma(\theta)}{\Sigma_{cr}},
\end{eqnarray}
where 
\begin{eqnarray}
\label{eq:surfacemassdensity_cr}
\Sigma_{cr} = \frac{c^2}{4\pi G}\frac{D_S}{D_LD_{LS}}
\end{eqnarray}
and $G$ is the gravitational constant, 
$c$ is the speed of light, and 
$D_L$, $D_S$, and $D_{LS}$ are the angular diameter distances to the lens object, to the background object, and from the lens to the background object, respectively.

\subsection{Point Mass}
The point mass model assumes that a mass $M$ is located only at the center of the lens object, so $\boldsymbol{\theta} = 0$ at the mass position. The lens potential, deflection angle, and distortions until 5th order for the point mass model are obtained as follows, assuming $\theta \neq 0$.
\begin{eqnarray}
\label{eq:pointmass_distortions}
\psi =& 2\ldst{0}{0} &=\theta_E^2 \ln(\theta)
\\
\bal =& 2\ldst{1}{1} &= \frac{\theta_E^2}{\bth^*} = \frac{\theta_E^2}{\theta}{\rm e}^{i\varphi}
\\
\cnv =& \ldst{2}{0} &= 0.0 
\\
\sh =& \ldst{2}{2} &=-\frac{\theta_E^2}{\bth^{*2}} = -\frac{\theta_E^2}{\theta^2}{\rm e}^{2i\varphi}
\\
\flf =& \ldst{3}{1} &= 0.0 
\\
\fls =& \ldst{3}{3} &=4 \frac{\theta_E^2}{\bth^{*3}} = 4\frac{\theta_E^2}{\theta^3}{\rm e}^{3i\varphi}
\\
&\ldst{4}{0} &= 0.0 
\\
&\ldst{4}{2} &= 0.0 
\\
&\ldst{4}{4} &=-24\frac{\theta_E^2}{\bth^{*4}} = -24\frac{\theta_E^2}{\theta^4}{\rm e}^{4i\varphi}
\\
&\ldst{5}{1} &= 0.0 
\\
&\ldst{5}{3} &= 0.0 
\\
&\ldst{5}{5} &= 192\frac{\theta_E^2}{\bth^{*5}} = 192\frac{\theta_E^2}{\theta^5}{\rm e}^{5i\varphi}
\hspace{5pt} 
\end{eqnarray}
where $\theta = |\bth|$ and $\theta_E = \sqrt{\frac{4GM}{c^2}\frac{D_{LS}}{D_LD_S}}$ is the Einstein radius.The general form can be obtained as follows
\begin{eqnarray}
\label{eq:pointmass_distortions_general}
\ldst{N}{M} &=&-(N-1)!(-2)^{N-2}\frac{\theta_E^2}{\theta^{*N}}  \delta_{NM}{\rm e}^{Mi\varphi} \hspace{5pt} (2 \leqq N \leqq 5)  .
\end{eqnarray}

\subsection{SIS}
The singular isothermal sphere (SIS) model has a three-dimensional mass density of $\rho(r) \propto r^{-2}$.
The lensing distortions of SIS are obtained as follows:
\begin{eqnarray}
\label{eq:SIS_distortions}
\psi =& 2\ldst{0}{0} &= \theta_E \theta
\\
\bal =& 2\ldst{1}{1} &= \theta_E {\rm e}^{i\varphi}
\\
\cnv =& \ldst{2}{0} &= \frac12\frac{\theta_E}{\theta}
\\
\sh =& \ldst{2}{2} &=-\frac12\frac{\theta_E}{\theta} {\rm e}^{2i\varphi}
\\
\flf =& \ldst{3}{1} &=-\frac12\frac{\theta_E}{\theta^2} {\rm e}^{i\varphi}
\\
\fls =& \ldst{3}{3} &=\frac32\frac{\theta_E}{\theta^2} {\rm e}^{3i\varphi}
\\
&\ldst{4}{0} &= \frac12\frac{\theta_E}{\theta^3}
\\
&\ldst{4}{2} &= \frac32\frac{\theta_E}{\theta^3} {\rm e}^{2i\varphi}
\\
&\ldst{4}{4} &=-\frac{15}2\frac{\theta_E}{\theta^3} {\rm e}^{4i\varphi}
\\
&\ldst{5}{1} &=-\frac32\frac{\theta_E}{\theta^4} {\rm e}^{i\varphi}
\\
&\ldst{5}{3} &=-\frac{15}2\frac{\theta_E}{\theta^4} {\rm e}^{3i\varphi}
\\
&\ldst{5}{5} &=\frac{105}2\frac{\theta_E}{\theta^4} {\rm e}^{5i\varphi}
\end{eqnarray}
where  $\theta_E = \frac{4\pi\sigma^2}{c^2}\frac{D_{LS}}{D_S}$ is Einstein radius.
Details of this model also can be found in Narayan (1997).
The general form can be obtained as follows
\begin{eqnarray}
\label{eq:SIS_distortions_general}
\ldst{N}{M} &=&\lr{-1}^N\lr{1-2\delta_{NM}}\frac{\lr{N+M-3}!!}2\frac{\theta_E}{\theta^{1-N}} {\rm e}^{Mi\varphi}\hspace{5pt} (2 \leqq N \leqq 5)
\end{eqnarray}
\subsection{Power Law}
The point mass model and the SIS model can be generalized as specific cases of a power-law model.  
This model is characterized by a mass density that falls off as a power law with index $l$, i.e.,
\begin{eqnarray}
\label{eq:powerlaw}
\rho(r) \propto r^{-l}.
\end{eqnarray}
In the limit $l \to 3$, the model reduces to the point-mass model, while for $l = 2$, it corresponds to the SIS model.  
The lensing parameters for this model can be calculated and are given by the following expressions.

\begin{eqnarray}
\label{eq:powerlaw_distortions}
\psi =& 2\ldst{0}{0} &=-\frac{\theta^2}{l-3}\lr{\frac{b}{\theta}}^{l-1}
\\
\bal =& 2\ldst{1}{1} &= \theta\lr{\frac{b}{\theta}}^{l-1}{\rm e}^{i\varphi}
\\
\cnv =& \ldst{2}{0} &=-\frac{l-3}{2}\lr{\frac{b}{\theta}}^{l-1}
\\
\sh =& \ldst{2}{2} &=-\frac{l-1}{2}\lr{\frac{b}{\theta}}^{l-1}{\rm e}^{2i\varphi}
\\
\flf =& \ldst{3}{1} &= \frac{\lr{l-3}\lr{l-1}}{2\theta}\lr{\frac{b}{\theta}}^{l-1}{\rm e}^{i\varphi}
\\
\fls =& \ldst{3}{3} &=\frac{\lr{l-1}\lr{l+1}}{2\theta}\lr{\frac{b}{\theta}}^{l-1}{\rm e}^{3i\varphi}
\\
&\ldst{4}{0} &=-\frac{\lr{l-3}\lr{l-1}\lr{l-1}}{2\theta^2}\lr{\frac{b}{\theta}}^{l-1}
\\
&\ldst{4}{2} &=-\frac{\lr{l-3}\lr{l-1}\lr{l+1}}{2\theta^2}\lr{\frac{b}{\theta}}^{l-1}{\rm e}^{2i\varphi}
\\
&\ldst{4}{4} &=-\frac{\lr{l-1}\lr{l+1}\lr{l+3}}{2\theta^2}\lr{\frac{b}{\theta}}^{l-1}{\rm e}^{4i\varphi}
\\
&\ldst{5}{1} &=\frac{\lr{l-3}\lr{l-1}\lr{l-1}\lr{l+1}}{2\theta^3}\lr{\frac{b}{\theta}}^{l-1}{\rm e}^{i\varphi}
\\
&\ldst{5}{3} &=\frac{\lr{l-3}\lr{l-1}\lr{l+1}\lr{l+3}}{2\theta^3}\lr{\frac{b}{\theta}}^{l-1}{\rm e}^{3i\varphi}
\\
&\ldst{5}{5} &=\frac{\lr{l-1}\lr{l+1}\lr{l+3}\lr{l+5}}{2\theta^3}\lr{\frac{b}{\theta}}^{l-1}{\rm e}^{5i\varphi}
\end{eqnarray}
The general form is 
\begin{eqnarray}
\label{eq:powerlaw_distortions_general}
\ldst{N}{M} &=& {\rm e}^{Mi\varphi}\lr{\frac{b}{\theta}}^{l-1}\lr{\frac{2}{\theta}}^{N-2}\lr{q-p^2}\prod_{j=1}^{N-2}\lr{q+p-j} \hspace{5pt} (2 \leqq N \leqq 5)
\end{eqnarray}
where $p=(N-M-2)/2$, $q=(3-l)/2$

\subsection{SIC}
The SIS model is an idealized and unphysical model because it has a divergent mass density at the center of the lens object.  
To solve this issue, various softened isothermal sphere models have been proposed.  
In this study, we adopt one such model: the softened isothermal sphere with core (SIC), as introduced in Narayan (1997).  
In this model, the lensing potential and deflection angle has a function of $\sqrt{\theta^2 + \theta_c^2}$ with core radius $\theta_c$,
and the deflection angle vanishes at the center of the lens object (i.e., at $\theta = 0$).

The lensing parameters of this model are calculated as follows,
\begin{eqnarray}
\label{eq:SIC_distortions}
\psi = &2\ldst{0}{0} &= \theta_E\theta \chi
\\
\bal = &2\ldst{1}{1} &= \theta_E \frac{1}{\chi}{\rm e}^{i\varphi}
\\
\cnv = &\ldst{2}{0} &= \frac12 \frac{\theta_E}{\theta} \frac{1+2x_c^2}{\chi^3}
\\
\sh = &\ldst{2}{2} &=-\frac12  \frac{1}{\chi^3}{\rm e}^{2i\varphi}
\\
\flf = &\ldst{3}{1} &=- \frac12 \frac{\theta_E}{\theta^2} \frac{1+4x_c^2}{\chi^5}{\rm e}^{i\varphi}
\\
\fls = &\ldst{3}{3} &= \frac32 \frac{\theta_E}{\theta^2} \frac{1}{\chi^5}{\rm e}^{3i\varphi}
\\
&\ldst{4}{0} &= \frac12 \frac{\theta_E}{\theta^3}\frac{1+8x_c^2-8x_c^4}{\chi^7}
\\
&\ldst{4}{2} &= \frac32 \frac{\theta_E}{\theta^3} \frac{1+6x_c^2}{\chi^7} {\rm e}^{2i\varphi}
\\
&\ldst{4}{4} &=- \frac{15}2 \frac{\theta_E}{\theta^3}\frac{1}{\chi^7}{\rm e}^{4i\varphi}
\\
&\ldst{5}{1} &=-\frac32 \frac{\theta_E}{\theta^4} \frac{1+12x_c^2-24x_c^4}{\chi^9}{\rm e}^{i\varphi}
\\
&\ldst{5}{3} &=-\frac{15}2 \frac{\theta_E}{\theta^4} \frac{1+8x_c^2}{\chi^9}{\rm e}^{3i\varphi}
\\
&\ldst{5}{5} &= \frac{105}2 \frac{\theta_E}{\theta^4} \frac{1}{\chi^9}{\rm e}^{5i\varphi}
\end{eqnarray}
the general form is 
\begin{eqnarray}
\label{eq:SIC_distortions_general}
\ldst{N}{M} &=& {\rm e}^{Mi\varphi}\frac{\theta_E}{\theta}\lr{\frac{2}{\theta}}^{N-2}\frac{1}{\chi^{2N-1}}\lr{\frac12-p^2}\prod_{j=1}^{N-2}\lr{\frac12+p-j} 
\nonumber\\
&&\times \lr{1+\frac{\lr{N+M}\lr{N-M}}{2}\theta^{'2}-\frac{\lr{N-M}\lr{N-M-2}\lr{N+M}\lr{N+M+2}}8\theta^{'4}}\hspace{5pt} (2 \leqq N \leqq 5)
\end{eqnarray}
where $\chi = \sqrt{1+x_c^2}$ and $x_c = \theta/\theta_c$

\subsection{NFW}
The Navarro–Frenk–White(NFW) model, introduced by Navarro, Frenk, and White (1995), describes the mass distribution of dark matter halos obtained from numerical simulations.  
It is characterized by a density profile given by 
\begin{eqnarray}
\label{eq:NFW}
\rho(r) \propto \frac{1}{(r/r_s)(1 + r/r_s)^2},
\end{eqnarray}
where $r_s$ is scaling radius of the NFW model.
The weak lensing flexion for the NFW model was calculated by Golse 2002, Bacon 2006.
In many previous studies, the surface mass density function of the NFW model is treated separately depending on the value of the real variable $x_s = \theta/r_s$. In this paper, however, we use
\begin{eqnarray}
\label{eq:NFW_function}
N_{00} &=& \frac{2}{\sqrt{x_s^2-1}}\arctan{\sqrt{\frac{x_s-1}{x_s+1}}} = \frac{2}{\sqrt{1-x_s^2}}\arctanh{\sqrt{\frac{1-x_s}{1+x_s}}},
\end{eqnarray}
based on the fact that the two functions coincide over the complex domain. While $N_{00}$ may take complex values during intermediate steps of the calculation, the final results are purely real.
Although the lensing parameters of this model are analytically complex, they can be calculated as follows.

\begin{eqnarray}
\label{eq:NFW_distortions}
2\ldst{0}{0} =&2k_sr_s^2\lr{\lr{\ln{\frac{x_s}{2}}}^2 + \lr{\sqrt{x_s^2-1}N_{00}}^2}&
\\
2\ldst{1}{1} =& 4k_sr_s\lr{\ln{\frac{x_s}{2}} + N_{00}}\frac{{\rm e}^{i\varphi}}{x_s} &\equiv \frac{4k_sr_s}{x_s}N_{11}{\rm e}^{i\varphi}
\\
\ldst{2}{0} =& 2k_s\frac{1 - N_{00}}{x_s^2 -1} &\equiv 2k_sN_{20} 
\\
\ldst{2}{2} =& 2k_s\lr{N_{20} - \frac{2}{x_s^2}N_{11}}{\rm e}^{2i\varphi} &\equiv  2k_sN_{22}{\rm e}^{2i\varphi}
\\
\ldst{3}{1} =& \frac{2k_sx_s}{r_s}\frac{1/x_s^2 - 3N_{20}}{x_s^2 -1}{\rm e}^{i\varphi} &\equiv \frac{2k_sx_s}{r_s}N_{31}{\rm e}^{i\varphi}
\\
\ldst{3}{3} =&\frac{2k_sx_s}{r_s}\lr{N_{31}-\frac{4}{x_s^2}N_{22}}{\rm e}^{3i\varphi}&\equiv \frac{2k_sx_s}{r_s}N_{33}{\rm e}^{3i\varphi}
\\
\ldst{4}{0} =&\frac{2k_s}{r_s^2x_s}\lr{N_{42}+\frac{2}{x^2}N_{31}}&\equiv \frac{2k_sx_s^2}{r_s^2}N_{40}
\\
\ldst{4}{2} =&\frac{2k_sx_s^2}{r_s^2}\lr{-\frac{2/x_s^4+5N_{31}}{x_s^2-1}}{\rm e}^{2i\varphi}&\equiv \frac{2k_sx_s^2}{r_s^2}N_{42}{\rm e}^{2i\varphi}
\\
\ldst{4}{4} =&\frac{2k_s}{r_s^2x_s}\lr{N_{42}-\frac{6}{x^2}N_{33}}{\rm e}^{4i\varphi}&\equiv \frac{2k_sx_s^2}{r_s^2}N_{44}{\rm e}^{4i\varphi}
\\
\ldst{5}{1} =&\frac{2k_sx_s^3}{r_s^3}\lr{N_{53}+\frac{4}{x_s^2}N_{42}}{\rm e}^{i\varphi} &\equiv \frac{2k_sx_s^3}{r_s^3}N_{51}{\rm e}^{i\varphi}
\\
\ldst{5}{3} =& \frac{2k_sx_s^3}{r_s^3}\lr{\frac{8/x_s^5 - 7N_{42}}{x_s^2-1}}{\rm e}^{3i\varphi} &\equiv \frac{2k_sx_s^3}{r_s^3}N_{53}{\rm e}^{3i\varphi}
\\
\ldst{5}{5} =& \frac{2k_sx_s^3}{r_s^3}\lr{N_{53}-\frac{8}{x_s^2}N_{44}}{\rm e}^{5i\varphi} &\equiv \frac{2k_sx_s^3}{r_s^3}N_{55}{\rm e}^{5i\varphi}
\end{eqnarray}
where $k_s$ is mass parameter of NFW model defined in Bartelmann 1996, and
\begin{eqnarray}
\frac{\partial}{\partial x_s}N_{11} &=& x_sN_{20}\\ 
\frac{\partial}{\partial x_s}N_{20} &=& x_sN_{31}\\ 
\frac{\partial}{\partial x_s}\lr{x_s^2N_{22}} &=& x_s^3N_{31}\\ 
\frac{\partial}{\partial x_s}N_{31} &=& x_sN_{42}\\ 
\frac{\partial}{\partial x_s}\lr{x_s^4N_{33}} &=& x_s^5N_{42}\\ 
\frac{\partial}{\partial x_s}N_{42} &=& x_sN_{53}\\ 
\frac{\partial}{\partial x_s}\lr{x_s^6N_{44}} &=& x_s^7N_{53}
\end{eqnarray}
and, each function \( N \) takes the following values in the limit \( x_s \to 1 \):
\begin{eqnarray}
\label{eq:NFW_distortions_N1}
N_{00} &=& 1\\
N_{11} &=& \ln{\frac12} + 1\\
N_{20} &=& \frac13\\
N_{22} &=& \frac13 -2\lr{\ln{\frac12} + 1}\\
N_{31} &=&-\frac25\\
N_{33} &=&-\frac25 + 4\lr{\frac13 -2\lr{\ln{\frac12} + 1}}\\
N_{40} &=& \frac87-\frac45\\
N_{42} &=& \frac87\\
N_{44} &=& \frac87+6\lr{\frac25 + 4\lr{\frac13 -2\lr{\ln{\frac12} + 1}}}\\
N_{51} &=&-\frac{48}9+\frac{32}7\\
N_{53} &=&-\frac{48}9\\
N_{55} &=&-\frac{48}9-8\lr{\frac87+6\lr{\frac25 + 4\lr{\frac13 -2\lr{\ln{\frac12} + 1}}}}.
\end{eqnarray}

\section{Investigation of Measurement Errors with Simulated Images}

In this section, we investigate the relationship between the maximum order of measured lensing distortions and the accuracy of reduced shear and reduced flexion estimates by generating simulated lensed images that contain all higher-order distortions. 

\subsection{Simulation}

In the simulation, we assume a circular source image with a Gaussian profile of radius $r_g$, and construct the lensed image by computing the lensing equation for each infinitesimal element of the background object, thereby including all higher-order distortions.  
The resulting lensed images are analyzed using the GEM.  
Because there is no intrinsic distortions in this simulation, the distortions we measure is lensing reduced distortions $\lrdst{N}{M}$, 
then the measured distortions ($\lrdst{N}{M}_{\rm meas}$) are then compared with the true input values ($\lrdst{N}{M}_{\rm true}$) to quantify the measurement errors.  

The error ratio $\lrdst{N}{M}_{\rm eratio}$ used in this section is defined as
\begin{eqnarray}
\label{eq:eratio}
\lrdst{N}{M}_{\rm eratio} = \left| \frac{\lrdst{N}{M}_{\rm meas} - \lrdst{N}{M}_{\rm true}}{\lrdst{N}{M}_{\rm true}} \right|.
\end{eqnarray}

We adopt several lens models, namely SIS, SIC, and NFW profiles, and investigate various values of lens strength, source position, and source size.  

\subsection{Relation between Maximum Order of Shape Measurement and Accuracy of Reduced Shear and Reduced Flexions}

\begin{figure}
 \begin{minipage}{0.495\hsize}
  \begin{center}
\vspace{0mm}
   \includegraphics[width=70mm]{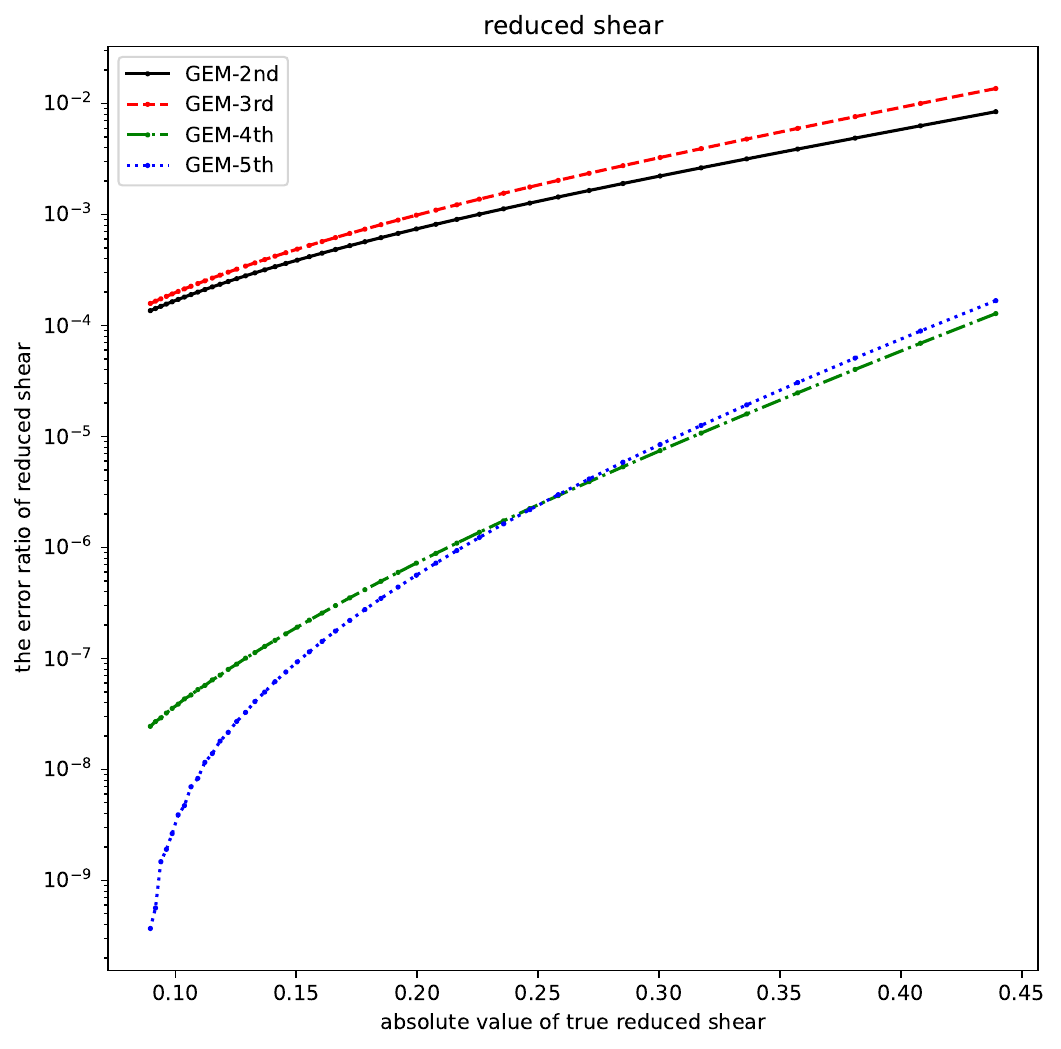}
  \end{center}
  \caption{
  Measurement error ratio in the reduced shear for a background object with $r_g=r_s/16$.
The horizontal axis shows the absolute value of lensing reduced shear.
The vertical axis shows the absolute error ratio in the measured reduced shear, calculated as the difference between the theoretical value from the lens model and the measured value.
The colors indicate the order of the shape measurement method.}
  \label{fig:Error_Shr_NFW_4}
 \end{minipage}
\hspace{3mm}
 \begin{minipage}{0.495\hsize}
  \begin{center}
  \vspace{0mm}
   \includegraphics[width=70mm]{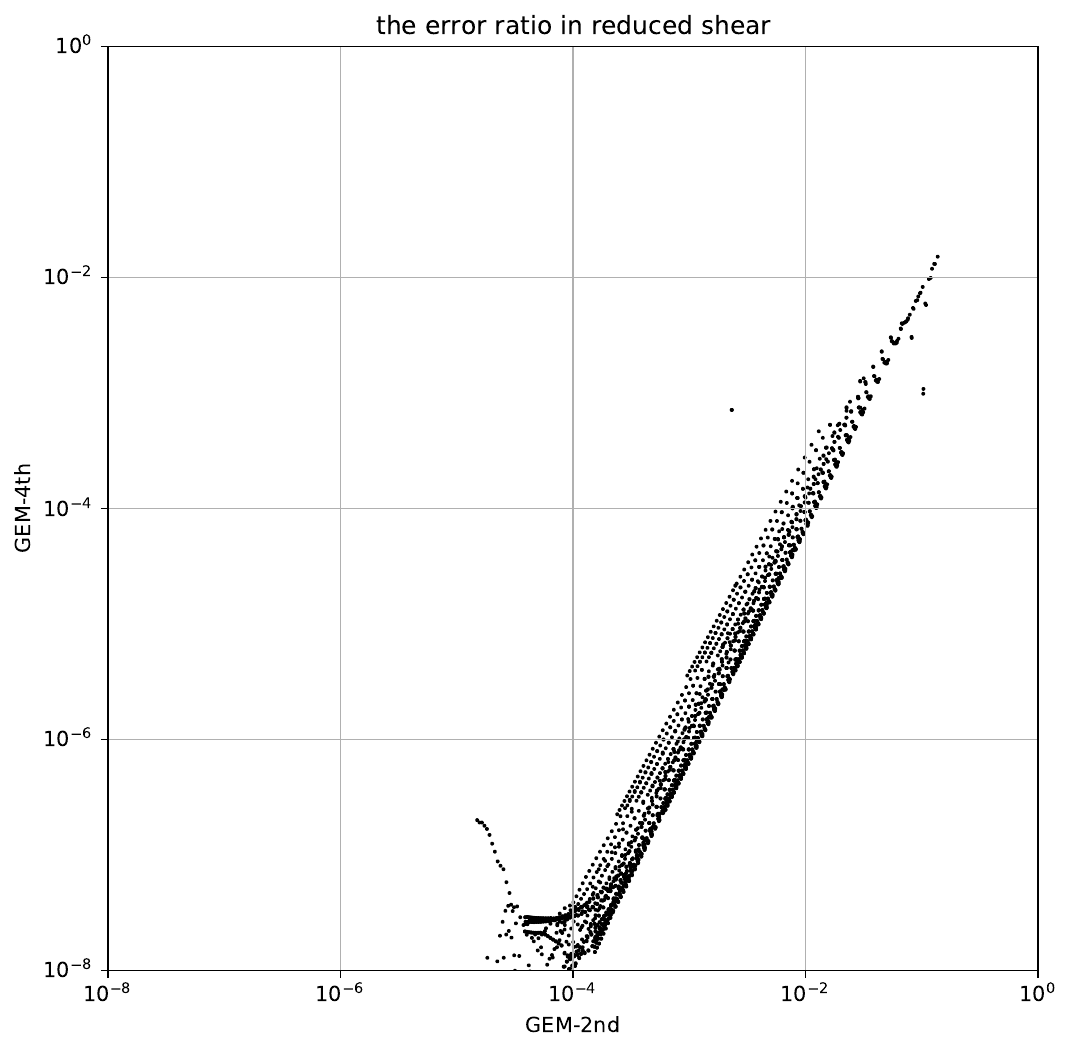}
  \end{center}
  \caption{
Comparison of the error ratios of reduced shear measured with GEM-2nd and GEM-4th.
The horizontal axis shows the error ratio of the reduced shear measured with GEM-2nd, and the vertical axis shows that measured with GEM-4th.\\
\\}
    \label{fig:Error_Shr_ALL}
 \end{minipage}
\end{figure}

Figure \ref{fig:Error_Shr_NFW_4} shows the error ratio of reduced shear for background sources  lensed by an NFW profile with $k_s=1$ and with size $r_g = r_s/16$ in the source plane, as a function of true reduced shear.  
While the GEM-2nd and GEM-3rd methods show situations where the error ratio exceeds $0.1\% \sim 1\%$, GEM-4th and GEM-5th show a significant improvement.  
The fact that the accuracy improves significantly when 4th-order terms are included indicates that the 4th-order distortion components are dominant contributor in this error. 
Figure \ref{fig:Error_Shr_ALL} compares the shear error ratios obtained with the GEM-2nd and GEM-4th for a variety of lens models, lens strengths, source positions, and source sizes.  
It is evident from this figure that including up to 4th-order terms consistently improves the shear error ratio across all situations.  
Residual errors even after including 4th-order terms are likely to originate from distortions of order higher than 4.  

\begin{figure}
 \begin{minipage}{0.495\hsize}
  \begin{center}
\vspace{0mm}
   \includegraphics[width=70mm]{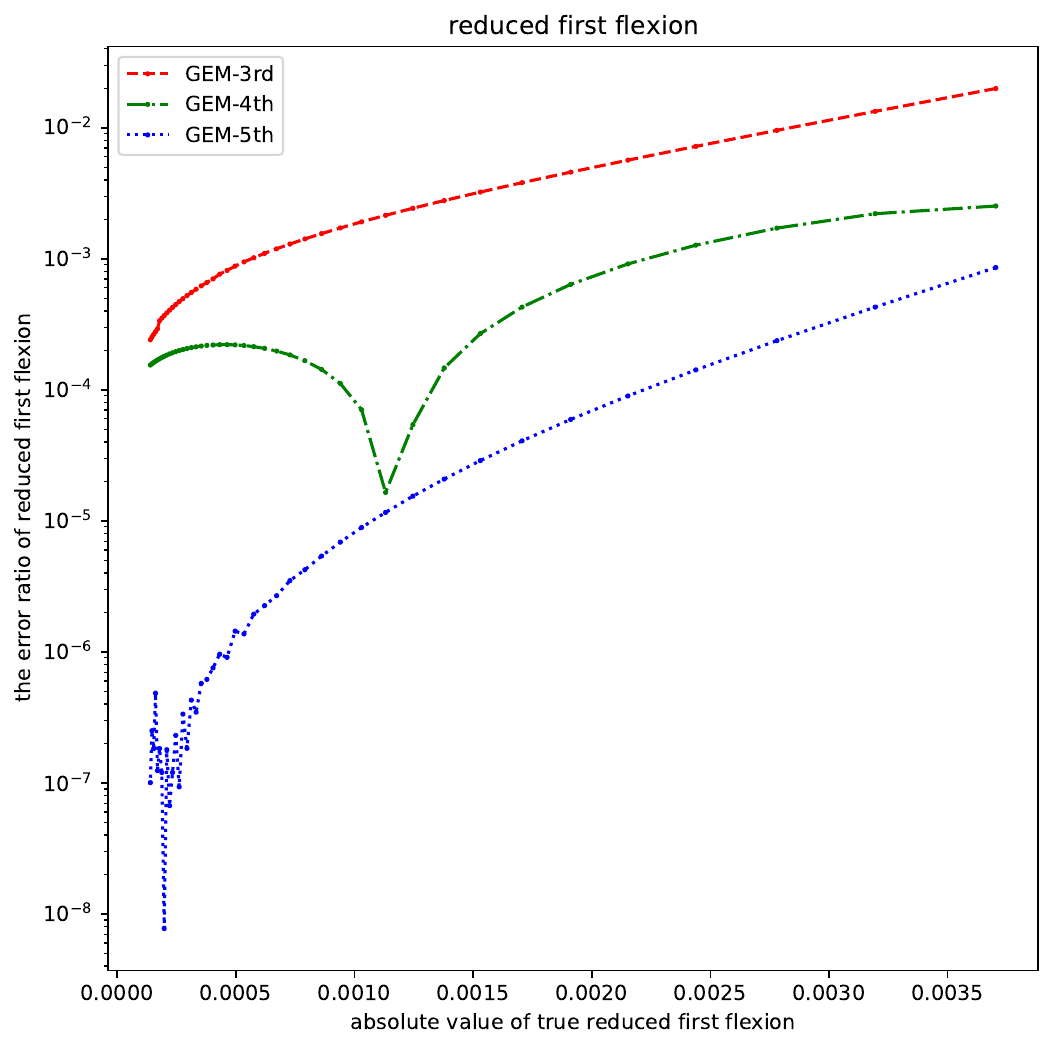}
  \end{center}
  \caption{Same as Figure \ref{fig:Error_Shr_NFW_4}, but showing the error ratio in the reduced first flexion.}
  \label{fig:Error_Fl1_NFW_4}
 \end{minipage}
\hspace{3mm}
 \begin{minipage}{0.495\hsize}
  \begin{center}
  \vspace{0mm}
   \includegraphics[width=70mm]{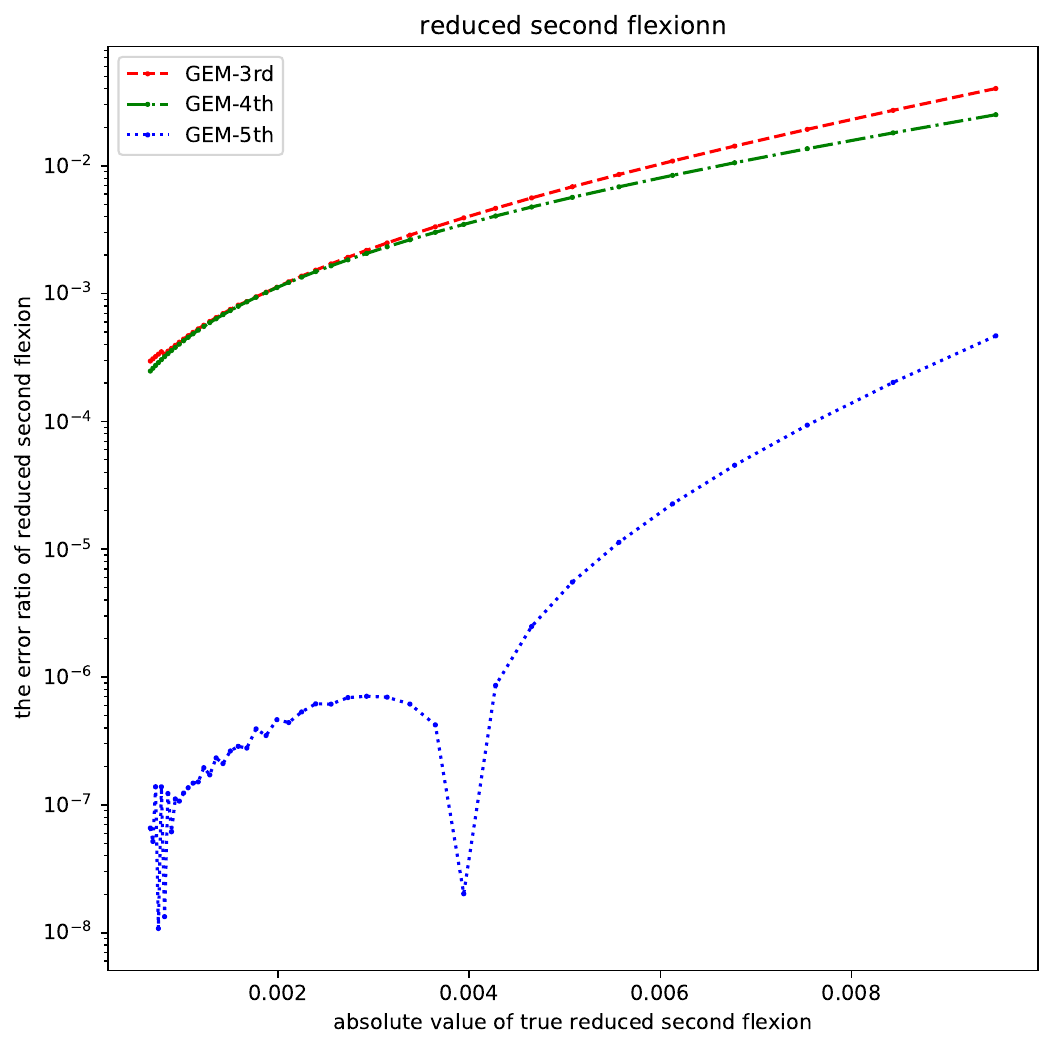}
  \end{center}
  \caption{Same as Figure \ref{fig:Error_Shr_NFW_4}, but showing the error ratio in the reduced second flexion.}
  \label{fig:Error_Fl2_NFW_4}
 \end{minipage}
\end{figure}

\begin{figure}
 \begin{minipage}{0.495\hsize}
  \begin{center}
\vspace{0mm}
   \includegraphics[width=70mm]{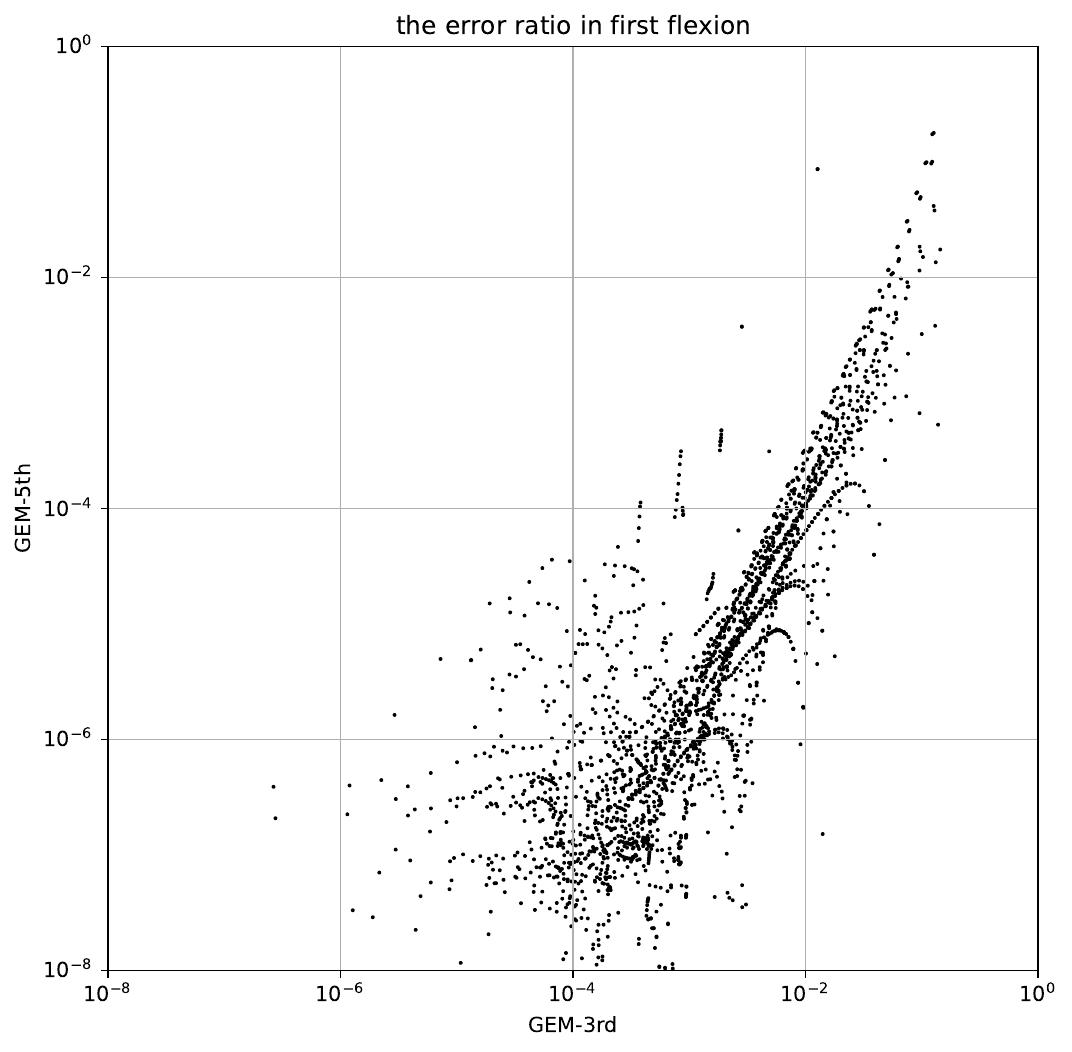}
  \end{center}
  \caption{Same as Figure \ref{fig:Error_Shr_ALL}, but showing the error ratio in the reduced first flexion measured GEM-3rd and GEM-5th.}
    \label{fig:Error_Fl1_ALL}
 \end{minipage}
\hspace{3mm}
 \begin{minipage}{0.495\hsize}
  \begin{center}
  \vspace{0mm}
   \includegraphics[width=70mm]{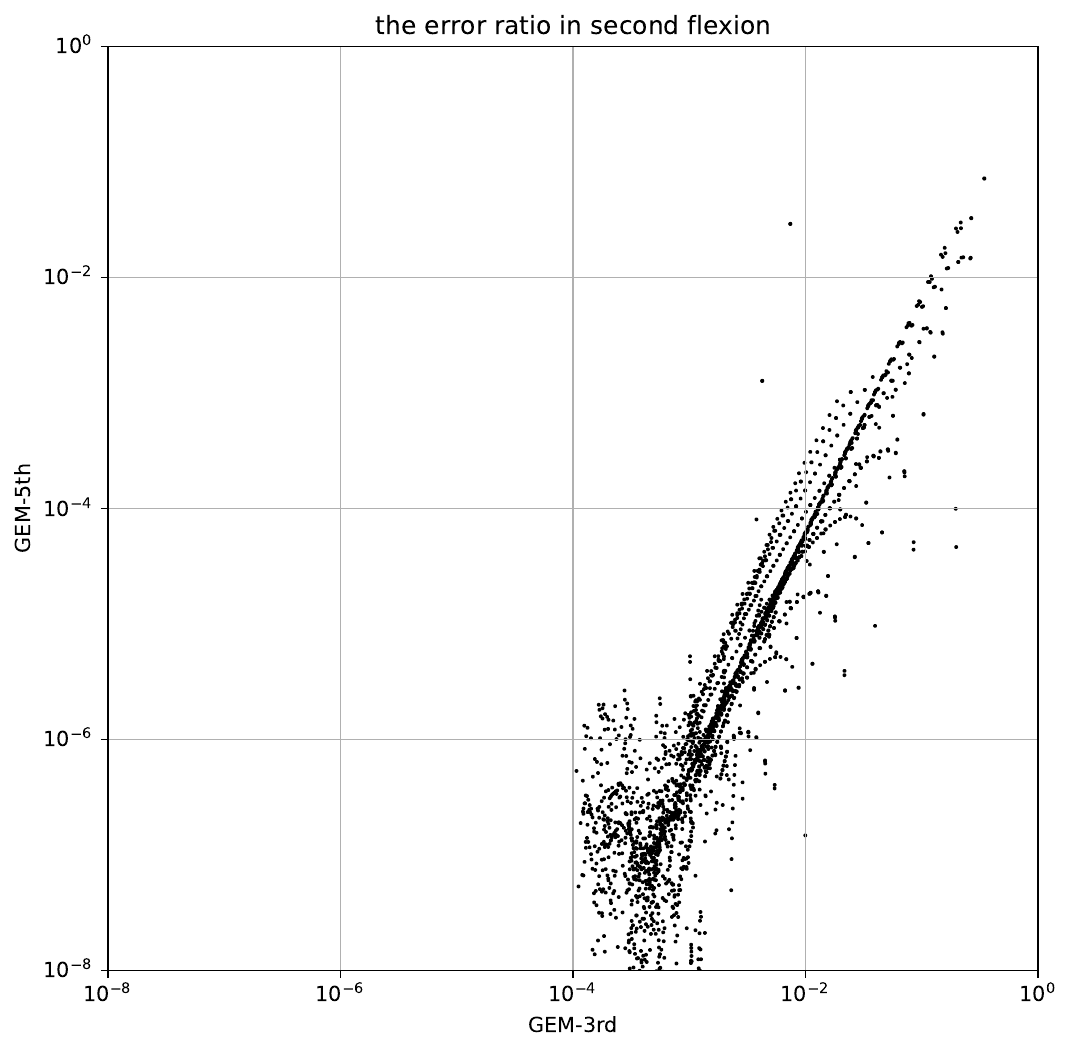}
  \end{center}
  \caption{Same as Figure \ref{fig:Error_Shr_ALL}, but showing the error ratio in the reduced second flexion measured GEM-3rd and GEM-5th.}
    \label{fig:Error_Fl2_ALL}
 \end{minipage}
\end{figure}

Figures \ref{fig:Error_Fl1_NFW_4} and \ref{fig:Error_Fl2_NFW_4} show, under the same conditions as in Figure \ref{fig:Error_Shr_NFW_4}, the error ratios of the first and second reduced flexions, respectively.  
These results indicate that flexion errors are strongly affected by 5th-order distortions, and that including terms up to 5th order yields substantial improvement.  
Figures \ref{fig:Error_Fl1_ALL} and \ref{fig:Error_Fl2_ALL} compare flexion error ratios between the GEM-3rd and GEM-5th for various lensing conditions, demonstrating that incorporating 5th-order terms consistently reduces errors.  


\begin{figure}
 \begin{minipage}{0.495\hsize}
  \begin{center}
\vspace{0mm}
   \includegraphics[width=70mm]{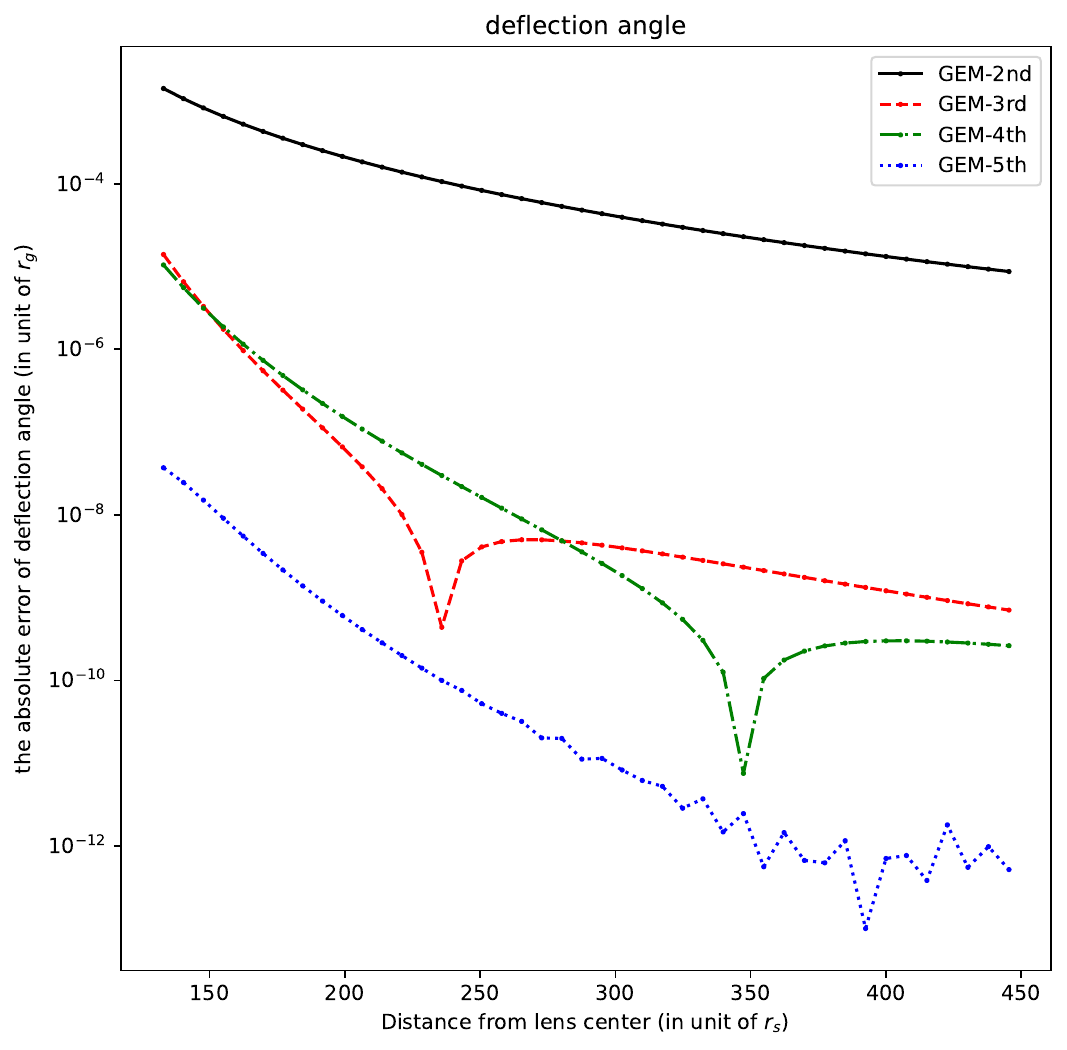}
  \end{center}
  \caption{Same as Figure \ref{fig:Error_Shr_NFW_4}, but showing the error (not error ratio) in the deflection angle.}
    \label{fig:Error_Dfa_NFW_4}
 \end{minipage}
\hspace{3mm}
 \begin{minipage}{0.495\hsize}
  \begin{center}
  \vspace{0mm}
   \includegraphics[width=70mm]{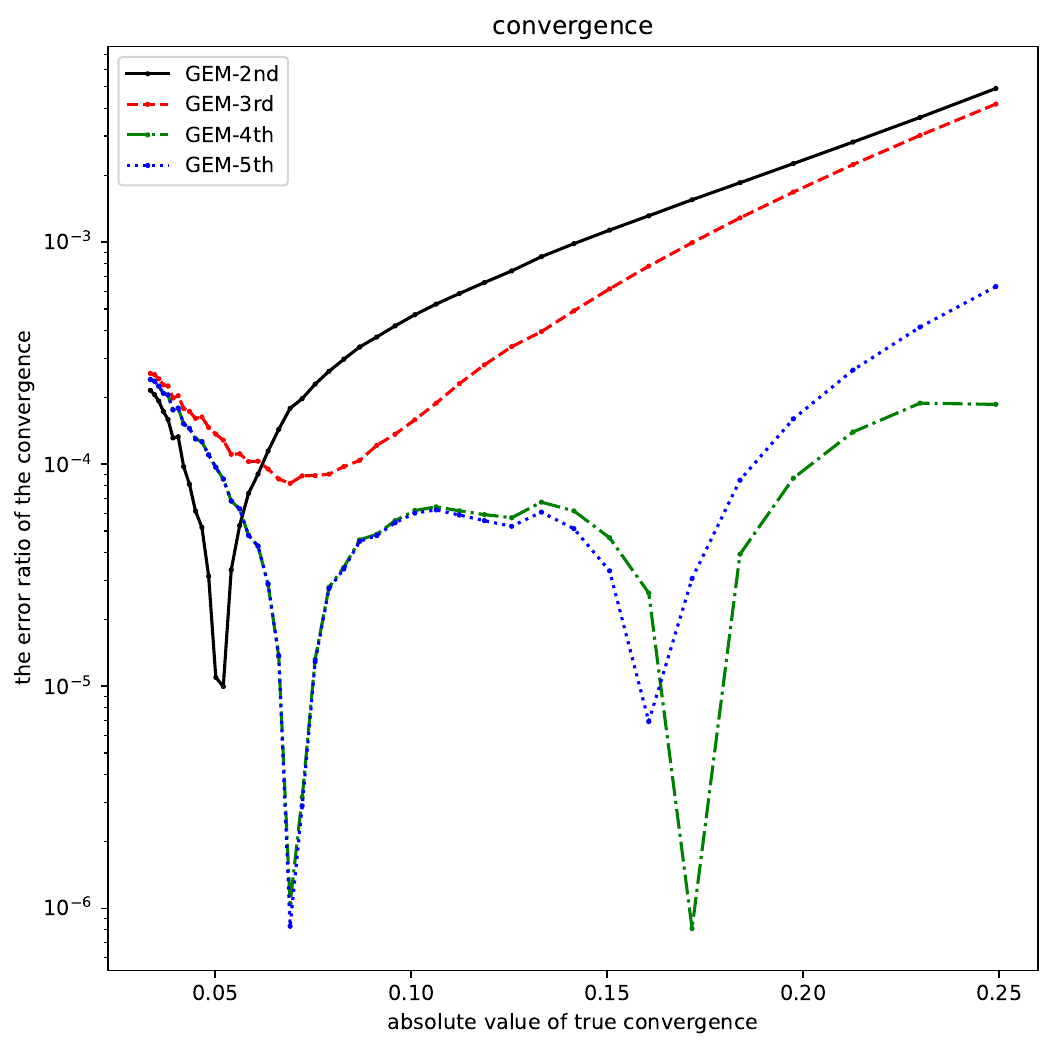}
  \end{center}
  \caption{Same as Figure \ref{fig:Error_Shr_NFW_4}, but showing the error ratio in the convergence derived from radius.}
  \label{fig:Error_Cnv_NFW_4}
 \end{minipage}
\end{figure}

Figures \ref{fig:Error_Dfa_NFW_4} and \ref{fig:Error_Cnv_NFW_4} present the error of the deflection angle and the error ratio of the convergence under the same conditions as Figure \ref{fig:Error_Shr_NFW_4}.  
Although these two quantities are not normally measured in weak lensing analyses, we show them here as a reference for future studies.  
The convergence is estimated from the difference in source-plane and image-plane radii as
\begin{eqnarray}
\label{eq:cnvrad}
\ldst{2}{0} = \frac{r_{\rm image}}{r_{\rm source}} - 1.
\end{eqnarray}
These figures show that the accuracy of the deflection angle can be greatly improved by including 3rd-order distortions with spin-1, while the accuracy of convergence can be enhanced by including 4th-order distortions with spin-0.  

\subsection{Radius dependency}
In the previous section, we showed that higher-order distortions introduce measurement errors. Since distortions of different orders have different dimensionalities in terms of length, their impact is expected to depend on the size of the background sources.
Therefore, in this section, we investigate how the impact of higher-order distortions changes when only the source size is varied while other conditions are kept fixed.

\begin{figure}
 \begin{minipage}{0.495\hsize}
  \begin{center}
\vspace{0mm}
   \includegraphics[width=70mm]{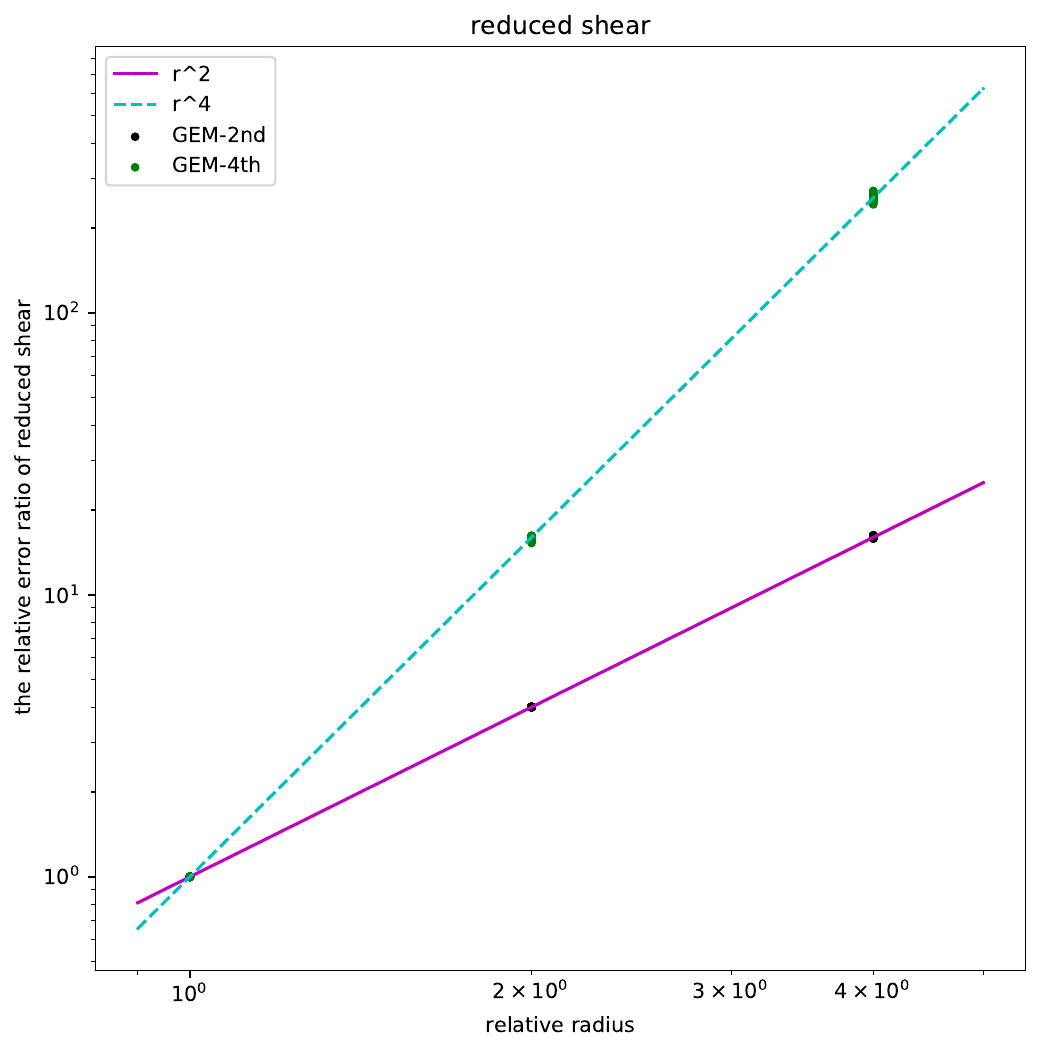}
  \end{center}
  \caption{Plot of the relative error ratio of the reduced shear as a function of the relative size of background sources.
The horizontal axis shows the relative size normalized by the fiducial background source size, and the vertical axis shows the relative error ratio normalized to unity at the fiducial size.
This illustrates how the error ratio changes when the background source size is doubled or quadrupled, while keeping all other parameters fixed.The solid purple line indicates the squared value of the relative radius with respect to the fiducial size, while the solid cyan line represents the fourth power.}
    \label{fig:Error_Shr_Rad}
 \end{minipage}
\hspace{3mm}
 \begin{minipage}{0.495\hsize}
  \begin{center}
  \vspace{0mm}
   \includegraphics[width=70mm]{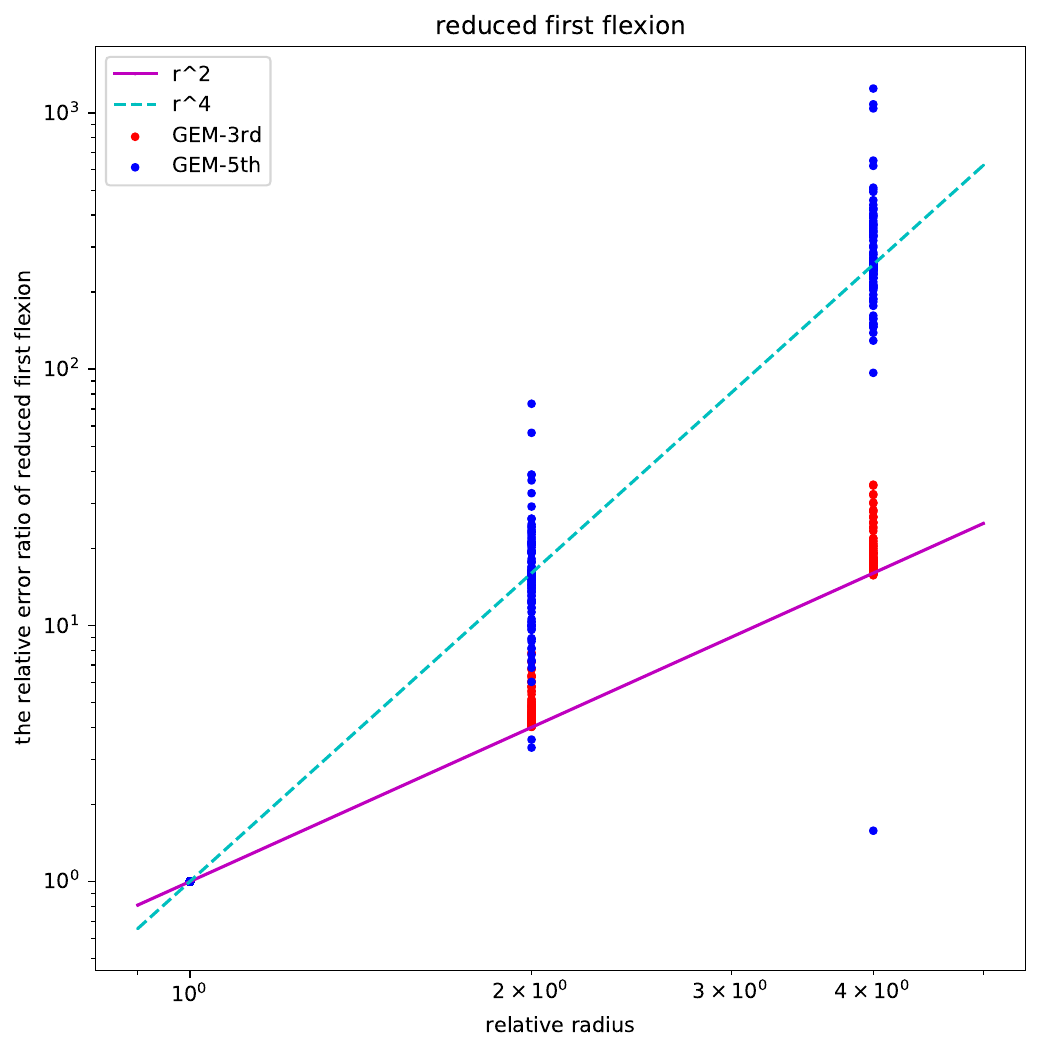}
  \end{center}
  \caption{Same figure as figure \ref{fig:Error_Shr_Rad}, but showing about the reduced first flexion.\\
  \\
  \\
  \\
  \\
  \\
  \\}
    \label{fig:Error_Fl1_Rad}
 \end{minipage}
\end{figure}

\begin{figure}
 \begin{minipage}{0.495\hsize}
  \begin{center}
\vspace{0mm}
   \includegraphics[width=70mm]{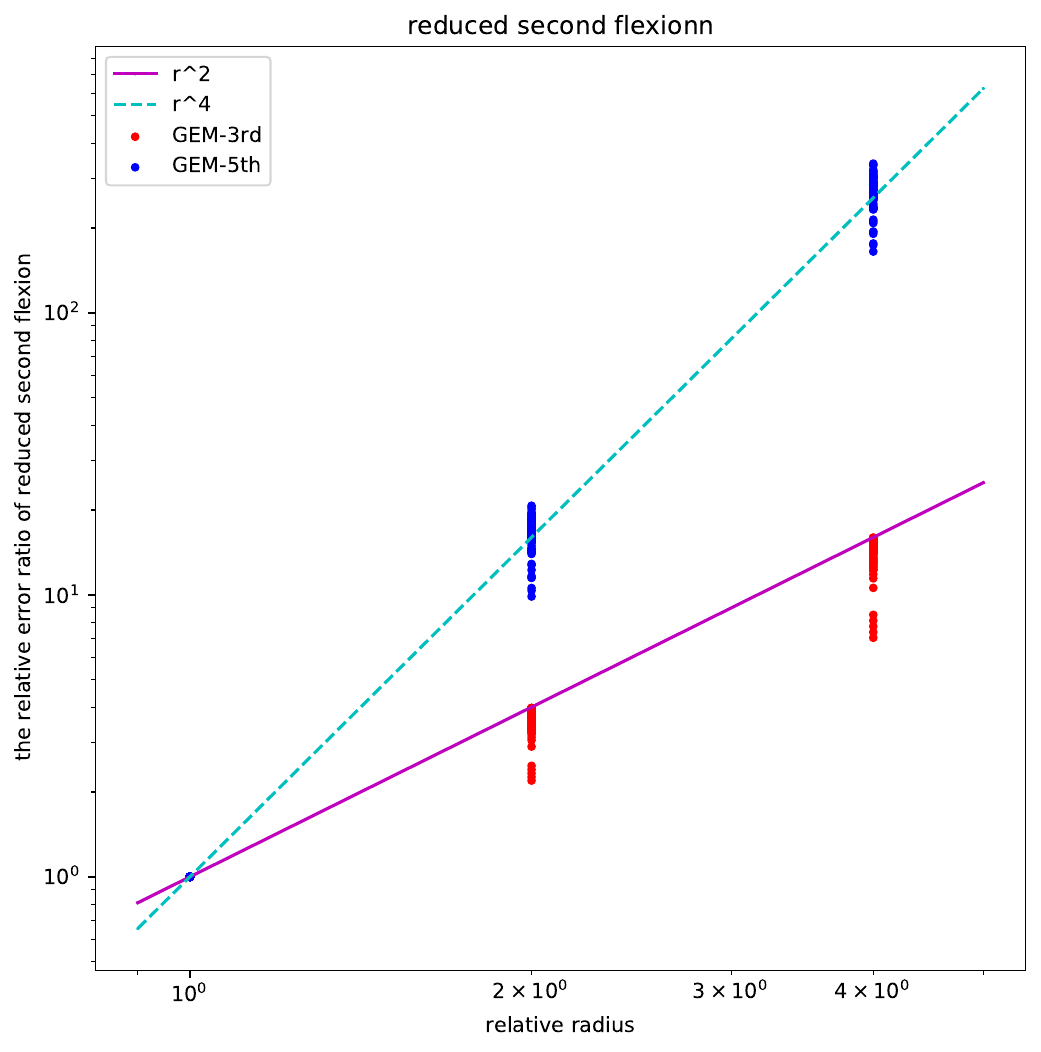}
  \end{center}
  \caption{Same figure as figure \ref{fig:Error_Shr_Rad}, but showing about the reduced second flexion.\\}
    \label{fig:Error_Fl2_Rad}
 \end{minipage}
\hspace{3mm}
 \begin{minipage}{0.495\hsize}
  \begin{center}
  \vspace{0mm}
   \includegraphics[width=70mm]{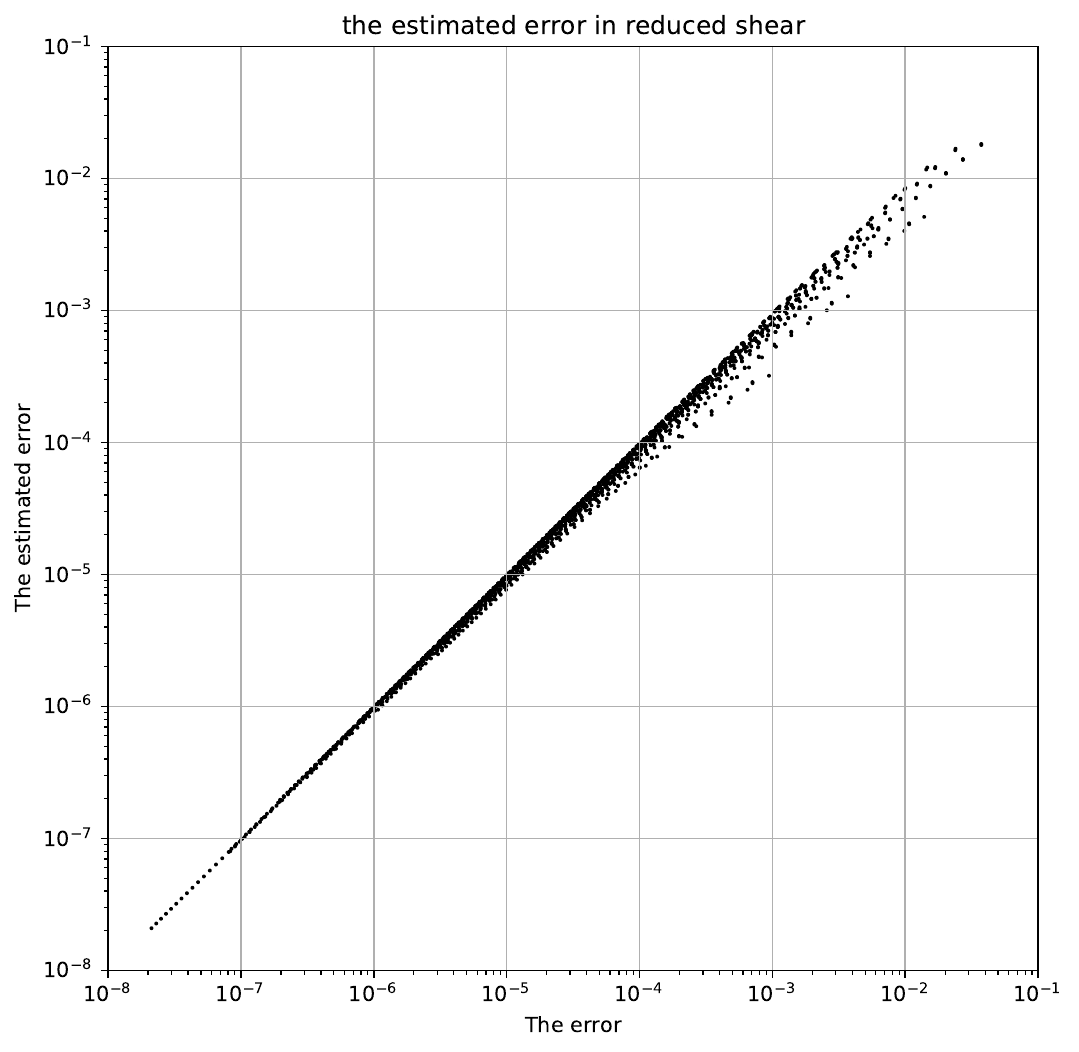}
  \end{center}
  \caption{Comparison with the error and the estimated error in the reduced shear}
    \label{fig:Error_Shr_Est}
 \end{minipage}
\end{figure}

Figure \ref{fig:Error_Shr_Rad} plots the error ratios of the reduced shear measured with GEM-2nd and GEM-4th as a function of relative source size. The results show a clear tendency that the error increases with the size of the background sources. Moreover, the error ratios measured with GEM-2nd are distributed along a quadratic curve, whereas those measured with GEM-4th are distributed along a quartic curve.
This indicates that, in the case of GEM-2nd, the dominant error originates from the fourth-order distortions, whose dimensionality differs by the square of length compared with shear. Similarly, in the case of GEM-4th, the dominant error arises from the sixth-order distortions, which differ by the fourth power of length.
Figures \ref{fig:Error_Fl1_Rad} and \ref{fig:Error_Fl2_Rad} present the results of similar tests for the first and second flexions. They show that when flexions are measured with GEM-3rd, the measurements are affected by higher-order distortions in a dependence on radius, similar to the case of shear.

From the results, we find that the impact of higher-order distortions becomes stronger for larger background sources and shows a strong correlation with the square of the source size.
In general weak lensing shape measurements, larger background objects are less affected by PSF effects, pixel noise, and pixelization when compared at the same surface brightness. However, our results indicate that the measurement errors induced by higher-order distortions are more significant for larger objects than for smaller ones.

\subsection{Estimating Errors from Higher-Order Distortions}
From the previous results, we conclude that the measurement of lensing distortions is dominantly affected by terms two orders higher than the measured quantity.  
Thus, measuring higher-order distortions makes it possible to remove their contaminating contributions and reduce measurement errors.  
On the other hand, in practical weak lensing analyses, the impact of higher-order distortions depends on the situation, and may be negligible in some cases, so higher-order measurements are not always required depending on the scientific goals.  
In this subsection, we describe a method to roughly estimate the strength of such higher-order contributions.  

\begin{figure}
 \begin{minipage}{0.495\hsize}
  \begin{center}
\vspace{0mm}
   \includegraphics[width=70mm]{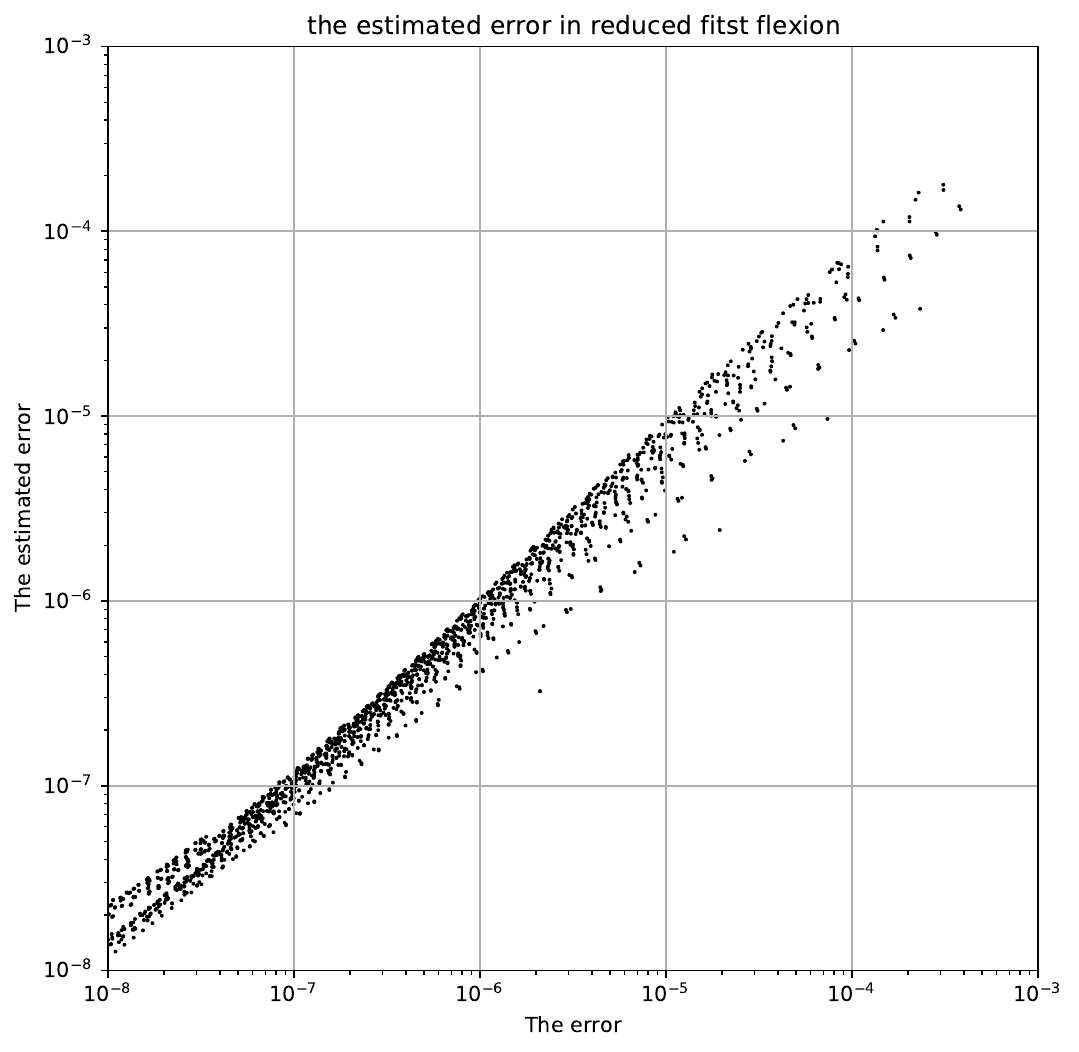}
  \end{center}
  \caption{Same figure as figure \ref{fig:Error_Shr_Est}, but showing the comparison about the reduced first flexion.}
    \label{fig:Error_Fl1_Est}
 \end{minipage}
\hspace{3mm}
 \begin{minipage}{0.495\hsize}
  \begin{center}
  \vspace{0mm}
   \includegraphics[width=70mm]{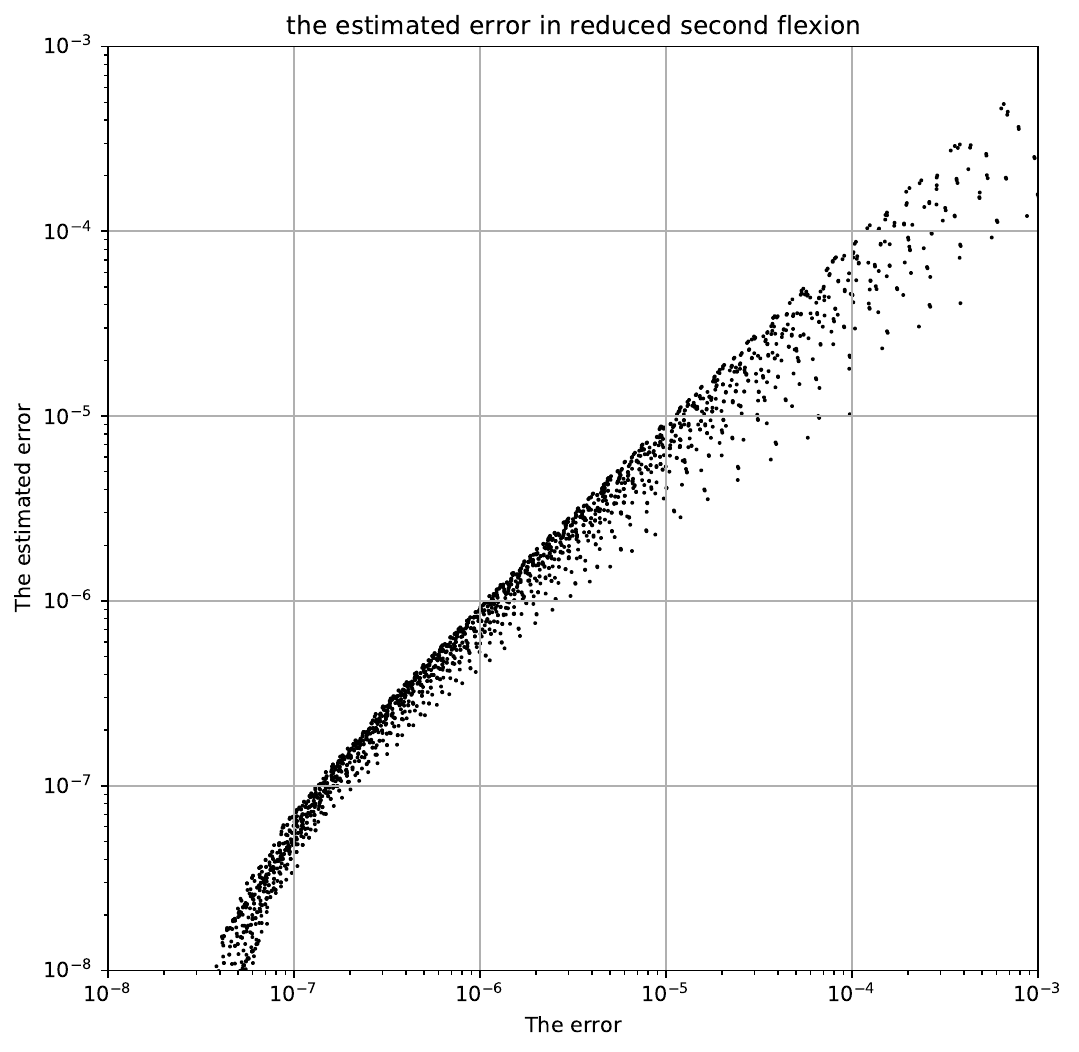}
  \end{center}
  \caption{Same figure as figure \ref{fig:Error_Shr_Est}, but showing the comparison about the reduced first flexion.}
    \label{fig:Error_Fl2_Est}
 \end{minipage}
\end{figure}

The error due to higher-order distortions can be approximately estimated as
\begin{eqnarray}
\label{eq:ErrorEstShr}
\lrdst{2}{2}_{estimated error} &=& \frac{1}{4}\frac{\zmom{4}{0}}{\zmom{2}{0}}\lrdst{4}{2},\\
\label{eq:ErrorEstFl1}
\lrdst{3}{1}_{estimated error} &=& \frac{11}{72}\frac{\zmom{6}{0}}{\zmom{4}{0}}\lrdst{5}{1},\\
\label{eq:ErrorEstFl2}
\lrdst{3}{3}_{estimated error} &=& \frac{10}{72}\frac{\zmom{6}{0}}{\zmom{4}{0}}\lrdst{5}{3}
\end{eqnarray}
Figures \ref{fig:Error_Shr_Est} to \ref{fig:Error_Fl2_Est} compare these estimates with the measured shear errors from the GEM-2nd and flexion errors from the GEM-3rd.  
The results show good agreement in most cases, confirming the validity of this rough estimation.  
Here, $Z^4_0/Z^2_0$ and $Z^6_0/Z^4_0$ are strongly correlated with the squared source size $r_g^2$, indicating that larger background objects are more affected by higher-order distortions.  
This result is consistent with the size dependence discussed in the previous section.

It should be noted, however, that in practical analyses the higher-order distortion terms ($\lrdst{4}{2}, \lrdst{5}{1}, \lrdst{5}{3}$) need to be either reconstructed from shear and flexion or obtained independently (e.g., through non-lensing methods) in advance.  

\section{Conclusions}
In this study, we investigated the impact of higher-order weak lensing distortions on the precise measurement of weak lensing shear and flexions, and developed methods to correct for these effects.
We began by extending conventional measurement methods, e.g. GEM-2nd and GEM-3rd, to include terms up to the fifth order in the expansion of the lens equation, thereby enabling the measurement of fifth-order weak lensing distortions.
This analysis framework allows flexible control over the maximum order of distortion to be measured. Using simulated images, we investigated how the measurement accuracy changes as a function of the highest order included.
In the simulation tests, we constructed lensed images by distorting circular source images using deflection angles computed from the several models. This approach ensures that the resulting images contain all orders of distortion as described by the full expansion of the lens equation.

The simulation results showed that, during the measurement of shear and flexions, higher-order distortions with the same spin number as the target quantity can be included as systematic errors. The amplitude of these errors increases with stronger lensing effects and larger background sources. In some cases, the error ratio exceeded 1\%.
However, these errors can be significantly reduced by directly measuring fourth- and fifth-order distortions. In the simulation tests, we were able to reduce the error ratios by one to three orders of magnitude.

In general, higher-order lensing distortions decrease more rapidly with distance than lower-order distortions. Therefore, in regions where the lensing effect is sufficiently weak, the impact of higher-order distortions is expected to be negligible. However, for lensing objects with significant mass fluctuations, such as sub-clumps within galaxy clusters, or for future cosmological analyses requiring extremely high precision, correcting for these higher-order effects may become essential.

Furthermore, if the intrinsic shapes of background galaxies are complex and include components that resemble higher-order distortions (e.g., ellipticities with radial dependence), these intrinsic higher-order shapes will contribute to the intrinsic noise in shear and flexion measurements. By measuring such higher-order shapes, it may be possible to disentangle intrinsic shear and flexion from higher-order intrinsic components, thereby reducing the overall intrinsic shape noise.


\section*{Data availability}
This paper uses only simulated images based on a simple lens model. The theoretical values of the weak lensing distortions derived from the lens model are also provided within this paper.





\end{document}